\documentclass[twocolumn,showpacs,preprintnumbers,superscriptaddress,prm,floatfix,aps,10pt]{revtex4-2}
\usepackage[utf8]{inputenc}
\usepackage{amsmath,amssymb}
\usepackage{xcolor}
\usepackage{todonotes}
\usepackage{comment}
\usepackage{xr-hyper}
\usepackage{url}
\usepackage{hyperref}
\hypersetup{
    colorlinks=true,
    citecolor=blue,
    linkcolor=blue,
    filecolor=magenta,      
    urlcolor=blue,
    }
\definecolor{olle}{RGB}{240,226,182}
\definecolor{nimrod}{RGB}{192,225,215}
\definecolor{omer}{RGB}{192,150,215}
\definecolor{maor}{RGB}{255,43,249}

\renewcommand {\vec}    [1]    {\ensuremath{\boldsymbol{#1}}}

\newcommand   {\frt}    [1]    {fourth-rank tensor}
\newcommand   {\srt}    [1]    {second-rank tensor}
\newcommand   {\Imm}    [1]    {\text{Im}}

\makeatletter

\newcommand*{\rom}[1]{\expandafter\@slowromancap\romannumeral #1@}

\makeatletter
\newcommand*{\addFileDependency}[1]{% argument=file name and extension
  \typeout{(#1)}
  \@addtofilelist{#1}
  \IfFileExists{#1}{}{\typeout{No file #1.}}
}
\makeatother

\newcommand*{\myexternaldocument}[1]{%
    \externaldocument{#1}%
    \addFileDependency{#1.tex}%
    \addFileDependency{#1.aux}%
}

\myexternaldocument{SI}

\begin{document}

%\title{Breakdown of mode assignment in the Raman spectrum of anharmonic crystals - Organic crystals as a case-study}
\title{Phonon-phonon interactions in the polarization dependence of Raman scattering}
\date{\today}

\author{Nimrod Benshalom}
\thanks{These authors contributed equally}
\affiliation{Department of Chemical and Biological Physics, Weizmann Institute of Science, Rehovot 76100, Israel}
\author{Maor Asher}
\thanks{These authors contributed equally}
\affiliation{Department of Chemical and Biological Physics, Weizmann Institute of Science, Rehovot 76100, Israel}
\author{R\'{e}my Jouclas}
\affiliation{Laboratoire de Chimie des Polym\`{e}res, Universi\'{t}e Libre de Bruxelles (ULB), 1050 Brussels, Belgium}
\author{Roman Korobko}
\affiliation{Department of Chemical and Biological Physics, Weizmann Institute of Science, Rehovot 76100, Israel}
\author{Guillaume Schweicher}
\affiliation{Laboratoire de Chimie des Polym\`{e}res, Universi\'{t}e Libre de Bruxelles (ULB), 1050 Brussels, Belgium}
\author{Jie Liu}
\affiliation{Laboratoire de Chimie des Polym\`{e}res, Universi\'{t}e Libre de Bruxelles (ULB), 1050 Brussels, Belgium}
\author{Yves Geerts}
\affiliation{Laboratoire de Chimie des Polym\`{e}res, Universi\'{t}e Libre de Bruxelles (ULB), 1050 Brussels, Belgium}
\affiliation{International Solvay Institutes for Physics and Chemistry, 1050 Brussels, Belgium}
\author{Olle Hellman}
\email{olle.hellman@weizmann.ac.il}
\affiliation{Department of Physics, Chemistry and Biology (IFM), Link\"oping University, SE-581 83, Link\"oping, Sweden}
\affiliation{Department of Molecular Chemistry and Material Science, Weizmann Institute of Science, Rehovot 76100, Israel}
\author{Omer Yaffe}
\email{omer.yaffe@weizmann.ac.il}
\affiliation{Department of Chemical and Biological Physics, Weizmann Institute of Science, Rehovot 76100, Israel}

\begin{abstract}
  We have found that the polarization dependence of Raman scattering in organic crystals at finite temperatures can only be described by a \frt\\ formalism. This generalization of the second-rank Raman tensor $\mathcal{R}$ stems from the effect of off-diagonal components in the crystal self-energy on the light scattering mechanism. We thus establish a novel manifestation of phonon-phonon interaction in inelastic light scattering, markedly separate from the better-known phonon lifetime. 
\end{abstract}

\maketitle

%\section*{TOC image}
%
% \begin{wrapfigure}{l}{0.55\textwidth}
%     \begin{center}
%         \includegraphics[width=0.8\linewidth]{TOC_image.png}    
%     \end{center}
% \end{wrapfigure}

%
%\begin{center}
%     \includegraphics[width=0.8\linewidth]{TOC_image.png}    
%\end{center}
%

\begin{figure*}[t!]
    \centering
    \includegraphics[width=\linewidth]{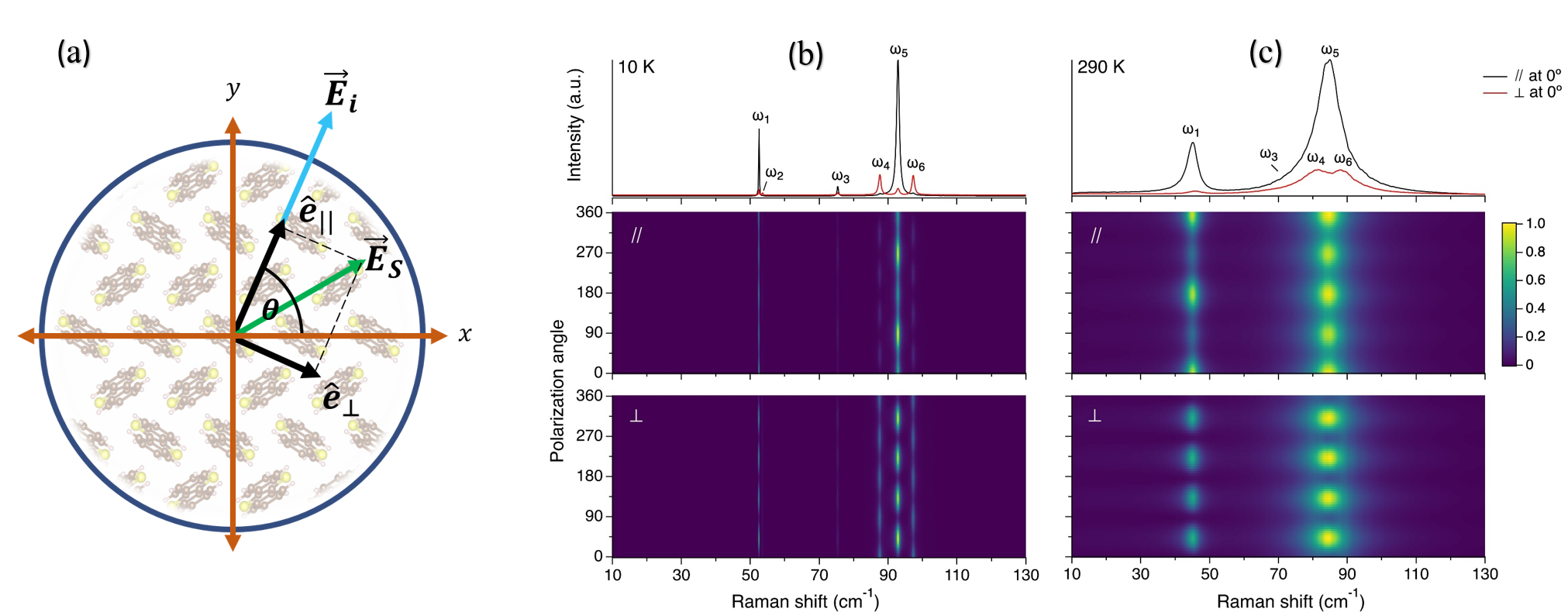}
    \caption{(\textbf{a}) Face-on schematic of a PO measurement. (\textbf{b}) PO measurement of BTBT at 10~K. (\textbf{c}) PO measurement of BTBT at 290~K. An example spectrum $(\theta=0^{\circ})$ for each configuration is given (top panels). Parallel measurement probes the scattered field's projection on $\hat{\vec{e}}_{\parallel}$ (middle panels), and cross measurement probes the scattered field's projection on $\hat{\vec{e}}_{\perp}$ (bottom panels).}
    \label{fig:setup}
\end{figure*}

The Raman spectrum of crystals is usually interpreted through the prism of mode assignment, where each spectral peak is assigned an irreducible representation (irrep) of the relevant space group~\cite{Hayes1978}.
So entrenched was this perspective, that past studies would denote their measurement geometry in terms of the irreps observable by it~\cite{Ganguly1972,Giintherodt1977,Montgomery1972,Potts1973,Haberkorn1973,Buchanan1974,Ishigame1987}.
This conceptual framework, also known as factor group analysis, provides a rigorous tool-set to connect observed spectra with crystal structure and vibrational symmetries.
Besides predicting the number of Raman active modes and their degeneracy, factor group analysis also dictates the polarization dependence of a spectral feature associated with a specific irrep.
The connection between the polarization of incident and scattered light is thus governed by the assigned irrep. 
Most textbooks~\cite{Yu2010,Long2002,Loudon1964,Pinczuk1975,Weber2000,Dresselhous2008}, as well as contemporary research~\cite{Livneh2006,Sander2012,Kranert2016,Sharma2020a}, use the matrix representation, also known as the Raman tensor, to describe this irrep-specific polarization dependence.
In this matrix formalism the scattering cross-section of a Raman peak, assigned an irrep $\Gamma_x$ and corresponding Raman tensor $\mathcal{R}_{\Gamma_x}$, is determined by~\cite{Yu2010}
\begin{equation}
 \sigma_{\Gamma_x}\propto\left|\hat{\vec{e}}_i\cdot\mathcal{R}_{\Gamma_x}\cdot \hat{\vec{e}}_s\right|^{2},
 \label{eq:cardona}
\end{equation}
where $\hat{\vec{e}}_i$ and $\hat{\vec{e}}_s$ are the incident, and collected light polarization unit vectors, respectively. 
This expression completely governs the polarization dependence of the Raman signal, with the angle entering through the geometrical relationship between the polarization unit vectors and crystal orientation (see Fig.~\ref{fig:setup}\textbf{a}).
The components of $\mathcal{R}_{\Gamma_x}$ are contracted susceptibility derivatives evaluated at equilibrium, so along with its persistent irrep identification Eq.~\eqref{eq:cardona} predicts a \emph{temperature independent} polarization pattern.

However, some of us recently showed the polarization-orientation (PO) Raman response, \textit{i.e.}, the polarization dependence of the Raman signal, evolves with temperature for some organic crystals~\cite{Asher2020,Asher2022}.
These kind of PO patterns and their temperature dependence can not be captured by any superposition of irrep specific matrices and a generalization to a fourth-rank formalism is necessary.
We argue this is the result of specific phonon-phonon interactions entering the Raman scattering process, an effect which is highly temperature and material dependent.
%We construct the appropriate \frt\\ by recognizing that as an observable, it must conform to all symmetry constraints of the crystal space group.
To further elucidate the physical origins of the effect, we adopt a generalized many-body treatment of the Raman scattering cross-section.
We find that by allowing off-diagonal components in the self-energy to persist, different irrep contributions in the spectrum are mixed, and a fourth-rank formalism is engendered.
Because the magnitude of these components is expected to be temperature dependent, this accounts for both the observed PO pattern and its temperature evolution. 
We confirm the validity of the model by successfully fitting it to experimental spectra.

This inadequacy of Eq.~\eqref{eq:cardona} stems from the na{\"i}ve treatment of temperature dependence in its derivation, which implicitly assumes a constant parabolic potential surface for atomic displacements, also known as the harmonic approximation~\cite{Dove2003,Yu2010}.
The role of anharmonicity in solid-state Raman spectroscopy has been the subject of extensive study, not only from an analytical point of view~\cite{Wallis1971,Loudon1963,Cowley1968}, but also from computational~\cite{Putrino2002,Raimbault2019,Benshalom2022}, and experimental~\cite{Spitzer1993,Widulle2001,Menedez1984,Mihailovic1993,Cardona2001} perspectives.
However, despite the consensus over the prominence of anharmonicity in Raman scattering, its effect on polarization has not been considered before \citet{Asher2020}.  

We apply our generalized model on the PO spectra of a few organic crystals, focusing on [1]benzothieno[3,2-$b$]benzothiophene (BTBT) as an ideal showcase.
As promising organic semiconducting molecular crystals, BTBT and its derivatives have been intensively investigated due to their high mobility and other application-oriented properties, such as air-stability and solution processability~\cite{Takimiya2006,Takimiya2014,Ebata2007,Schweicher2015,Wawrzinek2019,Ma2016}.
Besides its scientific importance, BTBT proved especially suitable for this study as its separate spectral features display the different possible PO trends with temperature, showing that the need for a fourth-rank formalism is not only material and temperature dependent, but may also be mode specific.

%\section{Polarization dependence of the Raman spectrum of BTBT}
%\subsection*{Results and discussion}
\label{BTBT_pol_Raman}

Our experimental setup employed a back-scattering geometry, as shown in Fig.~\ref{fig:setup}a.
In a PO measurement the polarization of the incident laser beam ($\vec{E}_i$) is rotated in a plane parallel to the BTBT (001) crystal face.
The inelastically back-scattered light can generally assume any polarization ($\vec{E}_s$). 
We separately probe the projection of scattered light onto either a parallel ($\hat{\vec{e}}_{\parallel}$) or cross ($\hat{\vec{e}}_{\perp}$) polarization with respect to the incident beam.
Full details for the measurement procedure and sample characterization are given in sections SI-II in the Supporting Information~\cite{Vyas2014,niebel2015,matsumura2016,Macrae2006,Macrae2008}.

Fig.~\ref{fig:setup}b and c present the Stokes Raman PO data for BTBT measured at 10~K and 290~K, respectively.
%Column (b) shows results for 10~K and column (b) for 290~K (RT).
At 10~K we observe exactly six sharp peaks, as predicted by Raman selection rules (Sec.~SIX.1 in the Supporting Information~\cite{Porto1981}).
%\ref{SI_BTBTPO_fit}
%This is an important validation for our subsequent analysis, as it makes the failure of the rigid-body approximation~\cite{Goldstein2002rigid} a less likely reason for inconsistencies. 
At 290~K we observe five distinguishable peaks, due to the low intensity of $\omega_2$ and the broadness of the spectrum.
The correspondence between spectral peaks across temperatures is corroborated by continuously sampling the Raman spectrum throughout the temperature range (Sec.~SVI in the Supporting Information).
%\ref{SI_BTBT_T_dependent}
%
\begin{figure*}[t]
    \centering
    \includegraphics[width=1.0\linewidth]{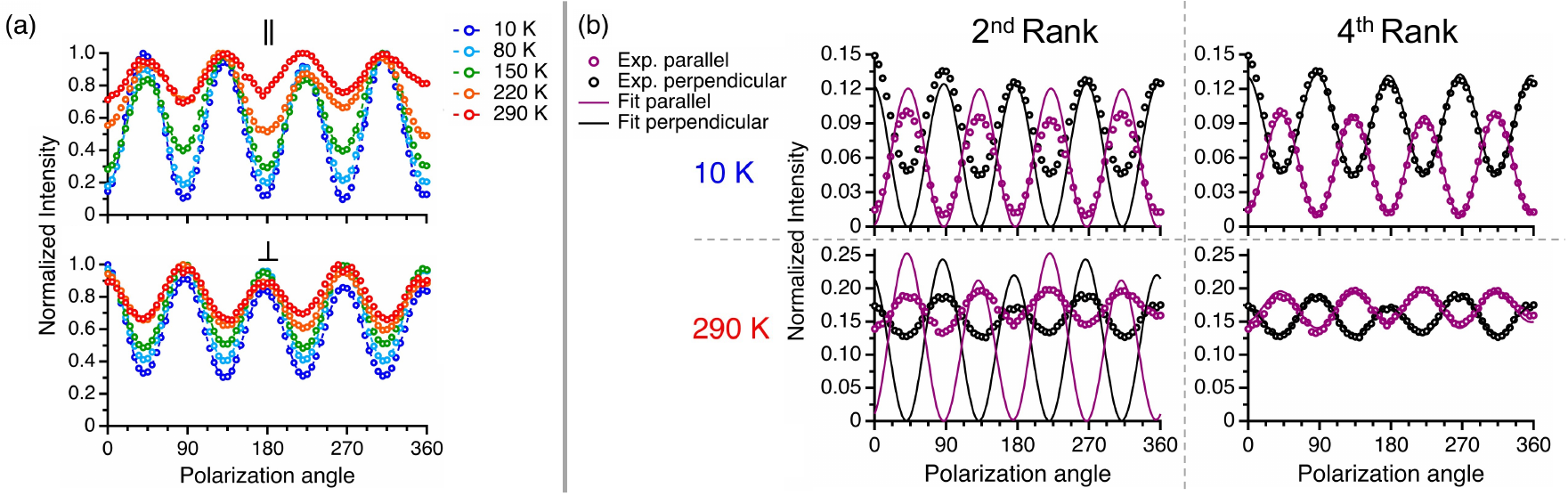}
    \caption{(a) The PO Raman response of $\omega_4$ from 10-290~K in parallel (top panel) and cross (bottom panel) configurations. (b) The fit results of the PO Raman response of $\omega_4$ at 10~K (top panels) and 290~K (bottom panels) using the second-rank (left panels), and fourth-rank Raman tensor formalism (right panels).}
    \label{fig:GoodFit_BadFit}
\end{figure*}
The polarization angle (vertical axis) in figures \ref{fig:setup}b and c is off-set by a random angle controlled by the crystal orientation in the laboratory frame, but is consistent for all spectra.

% Before examining the PO dependence of the spectra we must first associate a polarization dependent data series set, $\left\{I(\theta_i)\right\}$, with each spectral feature, in both parallel and cross configurations.
%The specifics of this step are to a large extent a matter of convenience.
%We use our PO Raman measurement to extract the polarization dependent integrated intensity of each spectral feature $(I(\theta))$ in both parallel and perpendicular configurations.
%We do so by fitting each spectrum to a multi-damped Lorentz line-shape, so the integrated intensity of each spectral feature may be directly evaluated (Sec.~S3 in SI~\cite{SI}).
%\ref{SI-fit}
To examine the PO dependence of each spectral feature we must first extract its integrated intensity.
We do this by fitting the raw spectra to a sum of six Lorentzians, in agreement with factor group analysis, and then extract the integrated intensity of each separate peak (see Sec.~SIII in the Supporting Information).
Because we are only interested in the polarization dependence, other deconvolution procedures may be chosen.  
We perform this process for the PO measurements at 10~K, 80~K, 150~K, 220~K, and 290~K (see sections SVII and SVIII in Supporting Information for the raw data and extracted temperature dependent PO pattern of each spectral feature, respectively).
%\ref{temp_PO_all_SI}\ref{T_dep_PO_SI}
With the PO pattern at hand, we can apply our model of choice to interpret it. %s polarization dependence.

Figure \ref{fig:GoodFit_BadFit}a presents the temperature evolution of the PO pattern for peak $\omega_4$ extracted from Fig.~\ref{fig:setup}, \textit{i.e.}, the change with temperature of the polarization dependence of its integrated intensity.
The PO pattern is highly temperature dependent in both parallel and cross configurations (see Sec.~SIV in Supporting Information for the temperature independent PO response of silicon~\cite{Lu2005}). %\ref{Si_PO_SI}
This temperature dependence in BTBT is fully reversible upon cooling, ruling out static disorder as its origin.

Following the harmonic approximation, we apply Eq.~\eqref{eq:cardona} and determine the appropriate irrep $\Gamma_x$ and its corresponding Raman matrix $\mathcal{R}_{\Gamma_x}$.
We account for birefringence effects by using a corrected Raman tensor, introducing a relative phase factor between the diagonal tensor components of $\mathcal{R}_{\Gamma_x}$~\cite{Kranert2016}.  
For our alternative, temperature dependent treatment, we follow \citet{Cowley1964} and \citet{Born1954} before him, who described the Raman cross-section with a \frt,
\begin{equation}
    \sigma(\Omega) \propto 
    \sum_{\mu\nu\xi\rho}
	n_{\mu}
	n_{\xi}
	I_{\mu\nu\xi\rho}(\Omega)
	E_{\nu}
	E_{\rho},
	\label{eq:frtf}
\end{equation}
where $\Omega$ is the probing frequency, $\vec{E}$ is the incident electric field and $\vec{n}$ is the unit vector for the observed polarization, with Greek letters standing for Cartesian components.
Because the Raman matrix in Eq.~\eqref{eq:cardona} is squared, casting it into a fourth-ranked tensor is immediately valid. 
However, its necessity compared to the prevalent second-rank formalism has never been demonstrated experimentally, nor was it considered theoretically.
We discuss the physical justification for Eq.~\eqref{eq:frtf} below, but for now treat $I_{\mu\nu\xi\rho}$ phenomenologically as the general material property governing light scattering in the crystal.
%For now, we treat $I_{\mu\nu\xi\rho}$ phenomenologically as the general material property governing the scattered light intensity, linearly induced in the crystal by an external incident field $\vec{E}$.
The tensor structure and component ratios of $I_{\mu\nu\xi\rho}$ will determine the PO pattern observed in experiment.
Unlike Eq.~\eqref{eq:cardona} this generalized form has the potential of producing non-negative values, but this is irrelevant when fitting experimental data to Eq.~\eqref{eq:frtf}.
We discusses the positive semi-definiteness of the scattering tensor in sec.~SIX.5 in the Supporting Information~\cite{nye1985,Slezak2020,VandenBos2007}.

Fig.~\ref{fig:GoodFit_BadFit}b shows the fitting attempts using both the predicted irrep (left panels), and generalized \frt\\ (right panels), for the same $\omega_4$ spectral feature at 10~K and 290~K (see sections SIX.2-3 in Supporting Information for all fitting tensors and results~\cite{Bilbao2006, Asher2020,Kranert2016,Yu2010}).
%\ref{SI_BTBTPO_fit}
%the fit results for the rest of the peaks at all measured temperatures using both models).
Using the harmonic Raman tensor, we find a partial fit at 10~K, which gradually deteriorates as temperature increases up to room temperature.
In stark contrast, the \frt\\ formalism gives a near perfect fit for all data sets.
A general \frt\\ obviously offers a much larger parameter space, but symmetry considerations dramatically reduce the number of independent variables in $I_{\mu\nu\xi\rho}$.  
We thus find the correct observable material property governing the Raman scattering around $\omega_4$ is a \emph{\frt\\.}
Similar behavior was previously observed in the Raman spectra of anthracene and pentacene crystals~\cite{Asher2020}, where again, the appropriate \frt\\ faithfully reproduces all PO patterns (see Sec.~SX in the Supporting Information~\cite{Asher2020}).

The matrix $\mathcal{R}$ used in Fig.~\ref{fig:GoodFit_BadFit}b is the specific irrep appropriate one.
However, even a completely unconstrained \srt\\, with an additional phase parameter for birefringence effects~\cite{Kranert2016}, was not able to successfully fit Eq.~\eqref{eq:cardona}.
Importantly, in our experiment this generalized form for $\mathcal{R}$ has more parameters than the symmetry reduced form of $I_{\mu\nu\xi\rho}$ (five vs.\ four).
%The explicit tensors and fitting results for both models are presented in sections SIX.2-3 in the SI~\cite{SI,Asher2020,Bilbao2006,Kranert2016,Yu2010}.
The different PO patterns made possible by each formalism depend on the measuring geometry, but one concrete example is demonstrated in Fig.~\ref{fig:GoodFit_BadFit};
in our back-scattering setup, any symmetric \srt\\ will make the cross signal vanish with 4-fold periodicity.
Indeed, this prohibits a successful fit to many of the observed spectra.
Conversely, the \frt\\ has no such limitation, allowing a faithful reproduction of the measured PO patterns.
%Next, we rationalize the use of a \frt\\ by considering the inelastic light scattering mechanism of anharmonic crystals.
%
\begin{figure*}[t!]
    \centering
    \includegraphics[width=\linewidth]{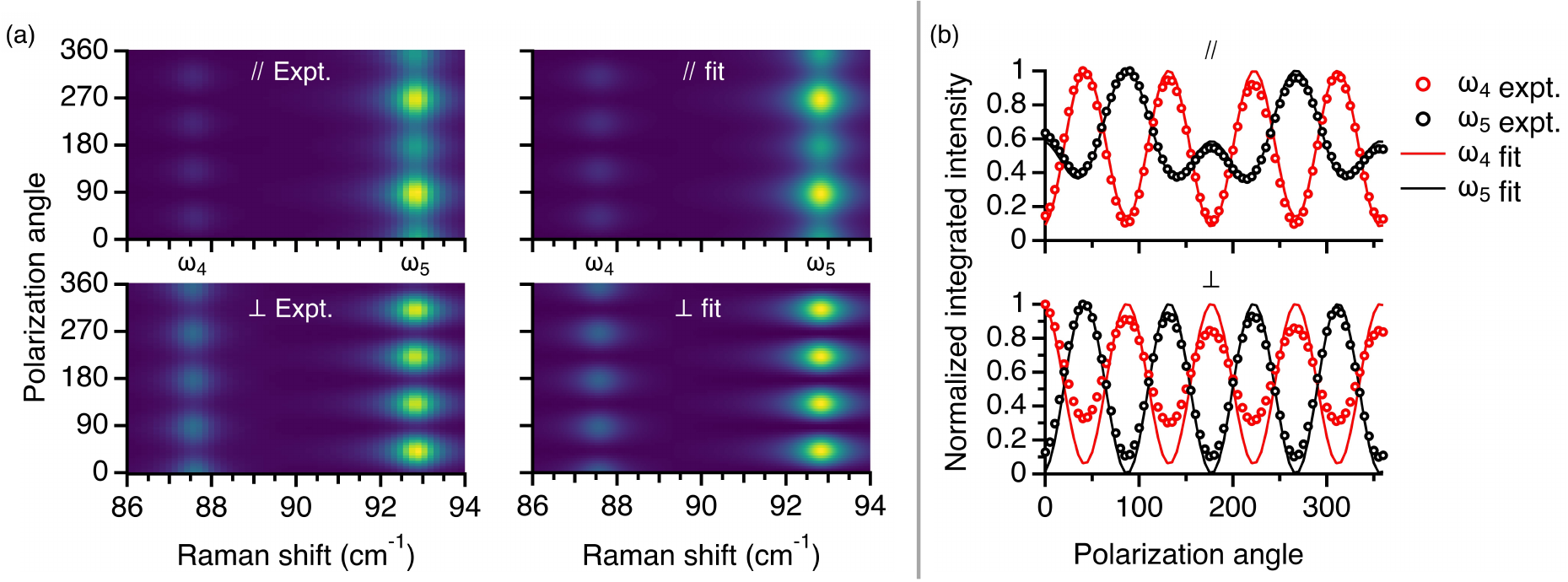}
    \caption{Global fit of the two-mode model for the PO patterns of spectral features $\omega_4$ and $\omega_5$ in the Raman spectra of BTBT. (a) Measurement (left) and fit (right) for parallel (top) and cross (bottom) configurations. (b) The integrated intensity of each feature as extracted from measurement (circles) and fit (lines).}
    \label{fig:2mode}
\end{figure*}

% \section{Polarization dependence of anthracene and pentacene}

% .......

%\section{The fourth-rank tensor description of Raman scattering}

%\subsection{Theory of inelastic light scattering at finite temperatures}

%We reiterate the tenets of the theory of inelastic light scattering from crystals at finite temperatures for clarity, and describe the novel aspect regarding polarization dependence demonstrated in this study.
We rationalize the use of a \frt\\ by considering the inelastic light scattering mechanism of anharmonic crystals.
Assuming only dipole scattering, the inelastic light scattering of a crystal as a function of probing frequency $\Omega$ is governed by~\cite{Cowley1964} 
\begin{equation}
    I_{\mu\nu\xi\rho}(\Omega,T)=\frac{1}{2\pi}\int\left\langle \chi_{\mu\nu}^{*}\left(t\right)\chi_{\xi\rho}\left(0\right)\right\rangle _{T}e^{-i\Omega t}\text{d}t,
    \label{eq:beast}
\end{equation}
the Fourier transform of the thermally averaged susceptibility auto-correlation function at temperature $T$.
The presentation here and full derivation in Sec.~SXI of the Supporting Information use Hartee atomic units~\cite{Leibfried1961,Cowley1963,wallace1998thermodynamics,Semwal1972,Benshalom2022,Dove2003,Kwok1968,Maradudin1962,Fultz2010}.
The fundamental physical entity - susceptibility ($\vec{\chi}$), is indeed a \srt\\ operator, but since experimentally we measure intensity, the observable is defined by the tensor product of two susceptibilities operators. 
Using normal mode coordinates $A_{\vec{q}s}$ for mode $s$ with momentum $\vec{q}$, and expanding the susceptibility in atomic displacements to first order, Eq.~\eqref{eq:beast} transforms to~\cite{Hayes19781st}
\begin{equation}
\label{eq:beast_part1}
    I_{\mu\nu\xi\rho}=\sum_{\lambda,\lambda^{'}}{\chi^{*}}^{\mu\nu}_{\lambda}\chi^{\xi\rho}_{\lambda'}
    \int
    \left<A_{\lambda}(t)A_{\lambda'}(0)\right>_{T}
    e^{-i\Omega t} \text{d}t,
\end{equation}
where we adopted a composite index $\lambda$ to denote $\vec{q}s$, and $\chi^{\mu\nu}_{\lambda}$ are the susceptibility derivatives with respect to normal-mode displacement $\lambda$. 
The thermal average $\left\langle\cdots\right\rangle_T$ in Eq.~\eqref{eq:beast_part1} is taken over crystal configurations. 
Except for the susceptibility derivatives, previously discussed by some of us~\cite{Benshalom2022}, the problem has reduced to one of lattice dynamics.

The ionic motion inside a crystal can be solved with the single-particle Green's function~\cite{Kwok1968}
\begin{equation}
    G_{\lambda\lambda'}(\Omega,T)= \mathrm{-}i\int\theta(t)\left<\left[A_{\lambda}(\tau),A_{\lambda'}(0)\right]\right>_{T}
     e^{\mathrm{-}i\Omega t} \text{d}t,
\end{equation}
where $\theta(t)$ is the Heaviside step function. 
It is commonly written as a sum of the harmonic Green's function $g^{0}_{\lambda\lambda'}$ and a complex self-energy term $\Sigma_{\lambda\lambda'}(\Omega)$:
\begin{equation}
\label{eq:Dyson}
    G(\Omega)^{-1} = g^0(\Omega)^{-1} + \Sigma(\Omega).
\end{equation}    
We can make Eq.~\eqref{eq:beast_part1} more analytically transparent by using the corresponding spectral function~\cite{Kwok1968},
\begin{equation}
\label{eq:Spectral}
J_{\lambda\lambda'}(\Omega,T)=-\frac{1}{\pi}\Imm\\\left\{G_{\lambda\lambda'}\right\}.
\end{equation}
The generalized anharmonic scattering tensor becomes   
\begin{equation}
\label{eq:general_I}
    I(\Omega,T)_{\mu\nu\xi\rho}=\sum_{\lambda\lambda'}{\chi^{*}}^{\mu\nu}_{\lambda}\chi^{\xi\rho}_{\lambda'}
    \left[n(\Omega,T)+1\right]
    J_{\lambda\lambda'}(\Omega),
\end{equation}
where $n(\Omega,T)$ is the Bose-Einstein occupation number for energy $\hbar\Omega$ at temperature $T$. 
%Next, we explain how expression~\eqref{eq:general_I} can give rise to an effective \frt\\ as well as account for its temperature evolution.

%\subsection{The two-mode model for inelastic light scattering}

To demonstrate how phonon-phonon interactions lead to a \frt\\ description, we evaluate Eq.~\eqref{eq:general_I} for the simplest case of two ``mixed'' normal modes with different irreps.
%spectral features $\omega_4$ and $\omega_5$ in BTBT (Fig.~\ref{fig:GoodFit_BadFit}).
%As detailed below, we allow for the minimal generalization required to account for the observed PO pattern.
The derivatives $\chi^{\mu\nu}_{\lambda'}$ are the same as in the harmonic treatment, so the Raman selection rules remain valid.
The BTBT structure only allows for A\textsubscript{g} and B\textsubscript{g} Raman active modes~\cite{Porto1981}.
%In that sense the same selection rules remain in effect.
However, these selection rules do not prohibit multiple terms contributing into the same frequency interval.
Retaining all Raman active modes will yield an intractable expression, so we consider the mixing of two modes with different irreps.
Absorbing the diagonal self-energy real part into the resonance frequencies $\omega_s$, our harmonic Green's function and self-energy term become~\cite{Kwok1968} 
\begin{equation}
    \label{eq:Green_matrices}
    g^{0}(\Omega)=
    \begin{pmatrix}
    \frac{2\omega_{\mathrm{A_g}}}{\Omega^{2}-\omega_{\mathrm{A_g}}^{2}} & 0\\
    0 & \frac{2\omega_{\mathrm{B_g}}}{\Omega^{2}-\omega_{\mathrm{B_g}}^{2}}
    \end{pmatrix};\quad
    \Sigma=\begin{pmatrix}i\Gamma_{\mathrm{A_g}} & \gamma\\
    \gamma & i\Gamma_{\mathrm{B_g}}
    \end{pmatrix},
\end{equation}
where we assumed real off-diagonal self-energy components $\gamma$, originally introduced by Barker and Hopfield to explain the infrared reflectivity of perovskites~\cite{Barker1964}, later used to explain their Raman spectra~\cite{Chaves1974,Sanjurjo1980}, but hitherto ignored in vector (polarization) analysis.
Importantly, these past studies invoked $\gamma$ to explain \emph{spectral line-shapes}, and are therefore open to competing diagonal self-energy descriptions.  
Because our PO measurements probe the \emph{tensor structure} of the scattering mechanism, we are able to exclude a diagonal model and thus demonstrate the unique signature of off-diagonal self-energy components.
Generally, the self-energy is also frequency dependent, $\Sigma=\Sigma(\Omega)$, but since the fitted spectral features are relatively sharp ($3~\mathrm{cm^{-1}}$), we assume a constant.
The full spectral function, (Eq.~S11), is given in Sec.~SXI.4 in the Supporting Information.

The mixed derivative terms $\vec{\chi}^{*}_{\mathrm{A_g}}\otimes\vec{\chi}_{\mathrm{B_g}}$ vanish due to symmetry considerations. 
Explicitly rewriting the tensorial part of Eq.~\eqref{eq:general_I} for our two-mode model, we obtain
\begin{equation}
\label{eq:2mode}
     \vec{I}(\Omega,T)\propto\vec{\chi}^{*}_{\mathrm{A_g}}\otimes\vec{\chi}_{\mathrm{A_g}}J_{\mathrm{A_gA_g}}+\vec{\chi}^{*}_{\mathrm{B_g}}\otimes\vec{\chi}_{\mathrm{B_g}}J_{\mathrm{B_gB_g}}.
\end{equation}
It would appear that we have recovered a quasi-harmonic expression, with each term corresponding to a separate spectral feature.
However, the off-diagonal component $\gamma$ dramatically changes the frequency dependence of the spectral functions $J(\Omega)$.
Instead of a single Lorentzian-like feature centered around a single resonance frequency, they now each have appreciable contributions in both $\omega_{\mathrm{A_g}}$ and $\omega_{\mathrm{B_g}}$.
It is impossible to re-create the tensorial structure of Eq.~\eqref{eq:2mode} by squaring a single matrix (Eq.~\eqref{eq:cardona}), and a \frt\\ formalism must be invoked.

Another way to understand the unique effect of the off-diagonal components is to examine the temperature dependence of the phonon's eigenvectors. 
Whereas diagonal self-energy components modify the phonon's lifetime and resonance frequency (\textit{i.e.}, eigenvalue), the off-diagonal components change the eigenvector, or identity, of the vibrational mode.
We can parameterize the linear combination describing the re-normalized eigenvectors using the self-energy term.
If the temperature-dependent  spectral function is diagonalized by
\begin{equation}
    \mathbb{J}(T) = Q^{-1} J(T) Q, 
\end{equation}
then the temperature dependent eigenvector $\vec{v}_{i}(T)$ is given by
\begin{equation}
    \vec{v}_{i}(T) = \sum_{j} \left(Q^{\mathrm{T}}\right)_{ij} \vec{v}^{0}_{j}, 
\end{equation}
where the index $j$ sums over all the zero-temperature (harmonic) eigenvectors $\vec{v}^{0}_j$ within the coupled subspace (two in our case), and the dimensionality of $\vec{v}_i$ is determined, as usual, by the number of atoms in the unit cell. 
% Although this only requires the diagonalization of the anharmonic spectral function, even in our simplified two-dimensional model, the resulting explicit dependency on the off-diagonal component $\gamma$ becomes extremely involved. 

The model described by equations \eqref{eq:frtf} and \eqref{eq:2mode} is tested by performing a global fit to the measured spectral range around $\omega_5$ (A\textsubscript{g}) and $\omega_4$ (B\textsubscript{g}) in BTBT (Fig.~\ref{fig:GoodFit_BadFit}).
Besides their appropriate irreps, our choice of peaks is guided by the PO pattern of $\omega_4$ that defies Eq.~\eqref{eq:cardona}, and by their spectral proximity, predicted to increase the effect of mixing between the spectral functions (see Eq.~(S11) in Sec.~SIX.4 in Supporting Information).  

The spectra for a full set of PO measurements (parallel and cross) at a given temperature ($T=10~K$) are fitted simultaneously to the two-mode model.
All fit functions, parameters and procedure are given in Sec.~SIX.4 in the Supporting Information.
Fig.~\ref{fig:2mode}a presents the zoomed-in PO measurement at 10~K alongside the fit results for the same frequency range.
The integrated intensity of each feature in each configuration was extracted from the fit and compared to the experimental values in Fig.~\ref{fig:2mode}b.
We find the parallel signal (top) matches the fit perfectly, with the cross (bottom) showing a slightly exaggerated oscillation amplitude, especially for $\omega_4$.

The global fit uses five parameters for all spectra, and is therefore not as accommodating as the fully general \frt\\ used in Fig.~\ref{fig:GoodFit_BadFit}, where the PO of \emph{each} peak is fitted to four independent parameters.
The pertinent comparison, however, is to the original harmonic model of Eq.~\eqref{eq:cardona}. 
Even when fully relaxing symmetry constraints and allowing for birefringence (totaling in seven independent parameters), Eq.~\eqref{eq:cardona} gave an inferior fit compared to the two-mode model.    
%Nonetheless, it is superior to the harmonic Eq.~\eqref{eq:cardona} (Fig.~\ref{fig:GoodFit_BadFit}(b)).
%From a physical point of view, we do not expect Eq.~\eqref{eq:2mode} to fully capture the PO spectra, which in reality rises from coupling of multiple modes. 
%Our simplified two-mode model is primarily meant to demonstrate how introducing the off-diagonal component $\gamma$ leads to an effective forth-rank description. 
We emphasize that although all equations are strictly rigorous, substantial simplifying approximations were introduced into the model.
%The analysis was limited to 1\textsuperscript{st}-order scattering, while higher-order scattering is expected to play a role in high temperatures.
Our choice to mix ``only'' two modes, as well as the choice of which two, is not uniquely prescribed by any physical principle, so we do not expect Eq.~\eqref{eq:2mode} to fit the PO spectra perfectly.
Furthermore, using a constant self-energy is at odds with the fully general frequency-dependent expression (Eq.~S29 in the Sec.~SXI.2 in the Supporting Information~\cite{Leibfried1961,Cowley1963,wallace1998thermodynamics,Semwal1972,Benshalom2022}). %\eqref{eq:self_energy}.
%Because of all of the above, limited utility should attributed to the extracted parameters.
We argue, however, that this approximated model is the minimally generalized treatment capable to explain the observed PO patterns and their temperature evolution.
Appreciable contributions from off-diagonal self-energy components is therefore a necessary qualification of any full description of Raman scattering in these systems.

\begin{figure*}[t]
    \centering
    \includegraphics[width=1.0\linewidth]{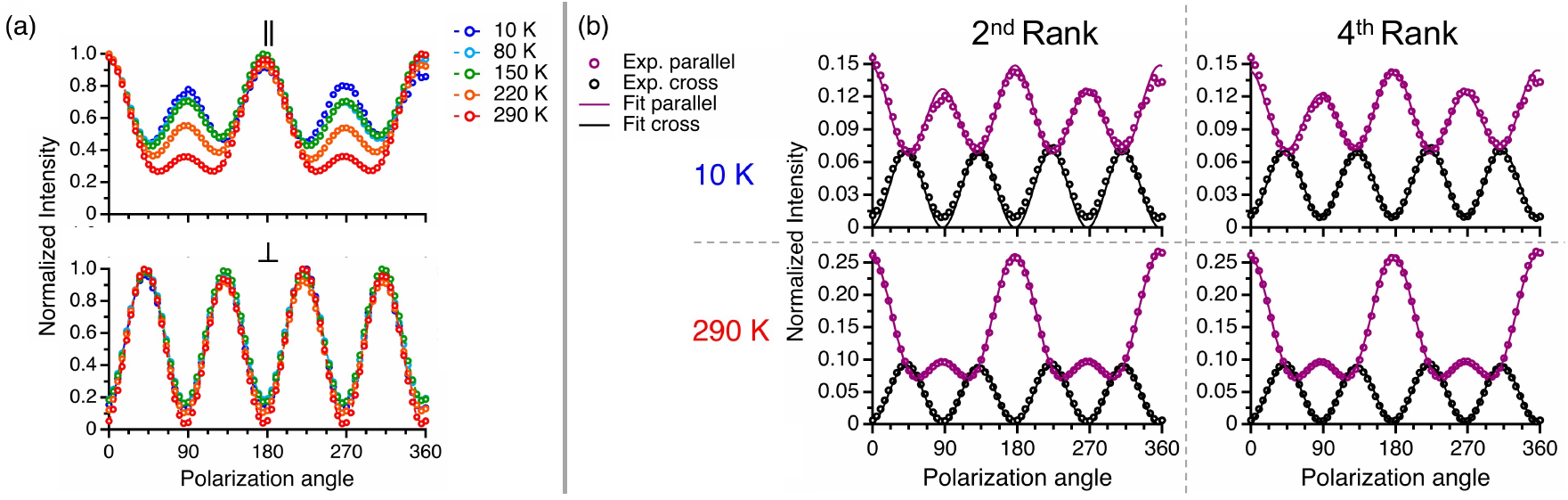}
    \caption{(a) The PO Raman response of $\omega_1$ from 10-290~K in parallel (top panel) and cross (bottom panel) configurations.
    (b) The fit results of the PO Raman response of $\omega_1$ at 10~K (top panels) and 290~K (bottom panels) using the second-rank Raman tensor formalism (left panels) and the fourth-rank Raman tensor formalism (right panels).}
    \label{fig:GoodFit_GoodFit}
\end{figure*}
%

%\subsection{Different manifestations of anharmonicity in light scattering}

The spectra and fits for another peak, $\omega_1$, are presented in Fig.~\ref{fig:GoodFit_GoodFit}.
Unlike the PO patterns of $\omega_4$, fitting Eq.~\eqref{eq:cardona} to $\omega_1$ proved satisfactory for all temperatures, with the generalized \frt\\ offering only a modest improvement (Fig.~\ref{fig:GoodFit_GoodFit}b).
Importantly, the PO pattern of $\omega_1$ still exhibits a clear evolution with temperature (Fig.~\ref{fig:GoodFit_GoodFit}a), which can not be explained by harmonic theory.

These two different scenarios demonstrate an important distinction in our interpretation of the mode mixing described by the self-energy off-diagonal components.
If an irrep-specific \srt\\ is able to capture the PO pattern, then same-irrep couplings have primarily contributed to the peak.
Because the anharmonic spectral function is temperature dependent, the weight of each contributing mode can change with temperature, altering the PO response.
Alternatively, whenever we observe a departure from the 4-fold cross pattern described above, %~(Sec.~\ref{BTBT_pol_Raman}), 
a \frt\\ must be invoked, and mixing between different irreps must be the culprit.
We can therefore immediately determine spectral feature $\omega_4$ (Fig.~\ref{fig:GoodFit_BadFit}) arises from appreciable $\left\langle\text{A}_g,\text{B}_g\right\rangle$ interactions, even without the benefit of an explicit fit.
This statement can potentially be tested further by numerical simulations~\cite{Raimbault2019}, or a 2D-Raman measurement~\cite{Hurwitz2020}, where inter-mode correlations are directly evaluated. 

Note that examination of any single BTBT spectrum would not hint at any extraordinary anharmonic effects, as the spectral features themselves are not particularly broad, nor do they display exceptional softening upon heating.
The significance of off-diagonal components is therefore not necessarily correlated with the overall magnitude of the self-energy term.
Indeed, it is perhaps the relative conventional spectrum shape that made the observation possible.

The two-mode model presented above allowed for a finite correlation between vibrational modes with different irreps.
Strictly speaking, these should vanish out of symmetry considerations~\cite{Maradudin1968}.  
This apparent contradiction is resolved by the fact that Raman scattering does not probe true $\Gamma$-point phonons.
Momentum conservation in the scattering process confers some finite momentum $\vec{q}$ to the vibrational excitation, determined by the measurement geometry and crystal orientation.
Both participating $\Gamma$-irreps map into identical irreps of $G\left(\vec{q}\right)$, the group of the wave-vector, so their coupling is allowed~\cite{Maradudin1968}. 
In other words, the Green's function remains invariant under the symmetry operations of the \vec{q}-vector and therefore need not vanish. 
In this sense, our use of the $\Gamma$-point susceptibility derivatives $\vec{\chi}_\Gamma$ is an approximation, retrospectively justified by the successful global fit.
%which changes the appropriate point-group symmetry to $G\left(\vec{q}\right)$, the group of the wave-vector \vec{q}.
%The specific identity of $\vec{q}$ is determined by the measurement geometry and crystal orientation.
% Specifically, our model proposed finite correlations between $\mathrm{A_g}$ and $\mathrm{B_g}$ modes.
% If we consider any $\vec{q}$-vector and track its compatibility table we'll find that both $\Gamma$-irreps necessarily map into the same irrep of $G\left(\vec{q}\right)$, so their coupling is allowed~\cite{Maradudin1968}.  

Finally, another possible generalization of the analysis presented is to consider light scattering from non-vibrational crystal excitations.
The distinction between the underlying dipole radiation mechanism, given in terms of a \srt\\, and the actual observable intensity, given in terms of a \frt\\, is completely general.
Considering the symmetry constrained \frt\\ as the fundamental material property might benefit other spectroscopic scenarios, such as exciton or magnon scattering.

%\section{Conclusions and Outlook}

In conclusion, we have demonstrated the unique role of anharmonicity in the polarization dependence of inelastic light scattering at finite temperatures.
We showed that the observed PO patterns of BTBT and other organic crystals can only be explained by generalizing the tensor structure governing Raman scattering to fourth-rank. 
This generalization is theoretically justified by allowing off-diagonal self-energy components to persist, which represents the mixing between vibrational modes and re-normalization of their eigenvectors.

The proposed fourth-rank model was realized by a sum of susceptibility tensor products with \emph{different} irreducible representations, and successfully tested by performing a global fit to the PO response of adjacent spectral features.
%Our observations were made possible by our unique measurement apparatus, which allows for a continuous PO measurement in a wide range of temperatures.

It is desirable to predict which vibrational modes will couple, as well as understand the temperature evolution of the off-diagonal self-energy components that describe them.
Perturbative methods offer a recipe for their calculation, given a known potential surface~\cite{Maradudin1962b,Cowley1963,Kwok1968}, but mapping chemical structure to such a surface is a formidable task in and of itself.
%Despite a comprehensive review of many modes across many materials~\cite{Asher2022}, we found no correlation between the observation of a temperature dependent PO pattern and the phonon symmetry, frequency or lifetime.
Generally, the effect is expected to depend on the whole of the vibrational dispersion relation, and require methods that provide the effective temperature dependent potential surface~\cite{Hellman2013,Benshalom2022}, to fully understand.   

%\section*{Author Contributions}
%We strongly encourage authors to include author contributions and recommend using \href{https://casrai.org/credit/}{CRediT} for standardised contribution descriptions. Please refer to our general \href{https://www.rsc.org/journals-books-databases/journal-authors-reviewers/author-responsibilities/}{author guidelines} for more information about authorship.

% \section*{Conflicts of interest}
% There are no conflicts to declare.

\section*{Acknowledgements}
O. Y. Acknowledges funding from the European Research Counsel (850041 - ANHARMONIC).
Support from the Swedish Research Council (VR) program 2020-04630 is gratefully acknowledged.
Y. G. is thankful to the Belgian National Fund for Scientific Research (FNRS) for financial support through research projects Pi-Fast (No T.0072.18), Pi-Chir (No T.0094.22), DIFFRA (No U.G001.19), 2D to 3D (No O.005018F), and CHISUB (No O.00322). Financial supports from the French Community of Belgian (ARC n◦ 20061) is also acknowledged. 
G. S. is a research associate of the FNRS.

%%%END OF MAIN TEXT%%%


%\balance


%%%REFERENCES%%%

\bibliography{main.bib} %You need to replace "rsc" on this line with the name of your .bib file

\end{document}

\begin{center}
    %\LARGE{\textit{Supplementary Material}}\\
    \LARGE{\textit{Supplemental Information}}\\
    \vspace{10pt}
    \large{\textbf{Phonon-phonon interactions in the polarizarion dependence of Raman scattering}}
\end{center}

\vspace{10pt}

%\noindent\large{\textbf{Phonon-phonon interactions in the polarizarion dependence of Raman scattering}}\\
\begin{center}
Nimrod Benshalom,$^{1,\ast}$ Maor Asher,$^{1,\ast}$ R\'{e}my Jouclas,$^{2}$ Roman Korobko,$^{1}$ Guillaume Schweicher,$^{2}$ Jie Liu$^{2}$, Yves Geerts,$^{2,3}$ Olle Hellman,$^{4,5,\dagger}$ and Omer Yaffe$^{1,\ddagger}$ \\
\textit{\small{
$^{1}$Department of Chemical and Biological Physics, Weizmann Institute of Science, Rehovot 76100, Israel.\\
$^{2}$Laboratoire de Chimie des Polym\`{e}res, Universi\'{t}e Libre de Bruxelles (ULB), 1050 Brussels, Belgium\\
$^{3}$International Solvay Institutes for Physics and Chemistry, 1050 Brussels, Belgium\\
$^{4}$Department of Physics, Chemistry and Biology (IFM), Link\"oping University, SE-581 83, Link\"oping, Sweden\\
$^{5}$Department of Molecular Chemistry and Material Science, Weizmann Institute of Science, Rehovot 76100, Israel
}}
\end{center}
~~~~~$\ast$ These authors contributed equally.

$\dagger$~\href{mailto:olle.hellman@liu.se}{olle.hellman@liu.se}

$\ddagger$~\href{mailto:omer.yaffe@weizmann.ac.il}{omer.yaffe@weizmann.ac.il} 

\vspace{20pt}

%Code to add "S" before sections and all that
\renewcommand{\thepage}{S\arabic{page}}  
\renewcommand{\thesection}{S\Roman{section}}   
\renewcommand{\thetable}{S\arabic{table}}   
\renewcommand{\thefigure}{S\arabic{figure}}
\renewcommand{\theequation}{S\arabic{equation}}

\setcounter{page}{1}

\large

\section{Measurement details}
\label{SI-experiment}

\textit{Crystals Growth and characterization:} [1]benzothieno[3,2-$b$]benzothiophene (BTBT) single-crystals were grown by leaving a saturated BTBT solution in chloroform in open air at room temperature until complete chloroform evaporation. 
We confirmed the crystal structure and high phase purity of the crystals by performing XRD measurements (see Sec.~\ref{SI-xrd}).

\textit{Temperature-Dependent PO Raman:} A custom-built Raman system was used to conduct the Raman measurements. 
The system included a 785~nm Toptica diode laser with intensity of around 30~mW on the sample.
To control the polarization of the incident and scattered light for the polarization-dependent measurements (5$^{\circ}$ steps), rotating half-wave plates and a polarizer-analyzer combination were used.
The system included a 50x objective.
Notch filters are included in the system to allow access to the low-frequency region ($>$10~cm$^{-1}$) and simultaneous acquisition of the Stokes and anti-Stokes signal. 
The system is based on a 1~m long Horiba FHR-1000 dispersive spectrometer with a 1800~mm$^{-1}$ grating. 
The spectral resolution was approximately 0.15~cm$^{-1}$.
The temperature was set and controlled by a Janis cryostat ST-500 and a temperature controller by Lakeshore model 335.

\clearpage

\section{\label{SI-xrd}Powder X-ray diffraction measurements of BTBT}

\begin{figure*}[h]
\centering
\includegraphics[scale=0.7]{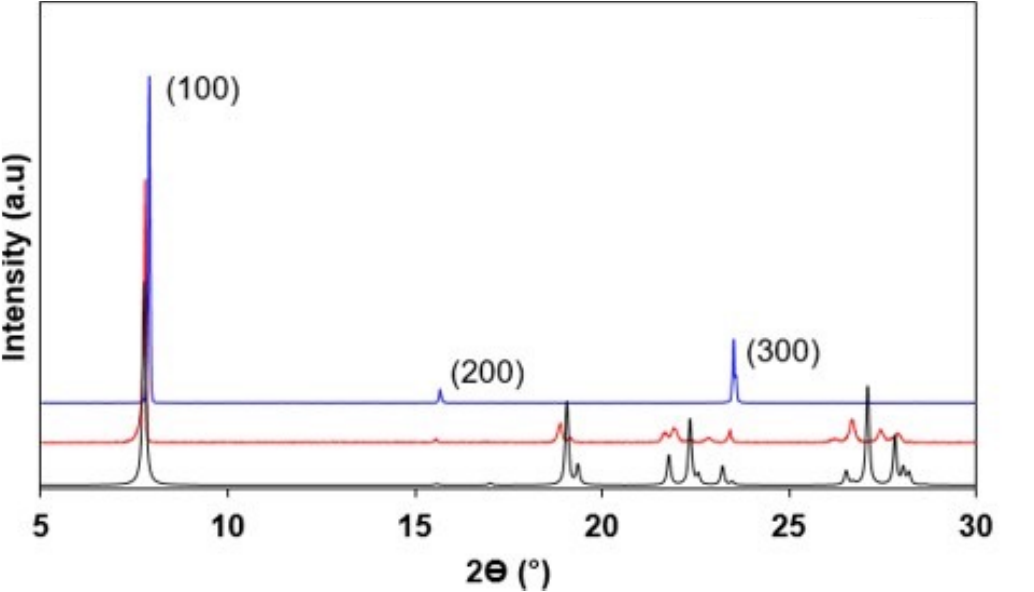}
\caption{\label{fig:XRD} X-Ray diffraction patterns calculated from the known crystal structures (in black), obtained experimentally from powders (in red) and obtained experimentally from single crystals (in blue) of BTBT.}
\end{figure*}

Preferred orientation and phase purity of BTBT were analyzed using powder X-ray diffraction. 
The experiment was performed at room temperature. 
The Compound was finely ground using a mortar and pestle for phase confirmation measurement. 
Single crystals were mounted over the sample holder for preferential orientation analysis.
For BTBT the measurements were conducted on a Panalytical Empyrean diffractometer using Cu–K$\alpha$ radiation ($\lambda=$1.54178~$\angstrom$). 
The diffractometer was set up with reflection-transmission spinner 3.0 configuration, and patterns were collected with $2\theta$ range between 5.0 and 30.0$^{\circ}$, steps of 0.1$^{\circ}$, time per step of 2.5~s, and rotation of 1~r/s. 
A calculated pattern was obtained from known crystal structure of BTBT~\cite{Vyas2014,niebel2015,matsumura2016} using Powder Pattern tool on Mercury software~\cite{Macrae2008,Macrae2006}.
The crystalline phase of BTBT was confirmed to have the same phase of the known crystal structures with CSD Refcode PODKEA02~\cite{matsumura2016} (BTBT) as it can be observed in Figure~\ref{fig:XRD} (experimental diffraction pattern in red and calculated diffraction patterns in black). 
The results show single crystals of BTBT with highly (100) preferred orientation crystallized along [001] direction.

\clearpage

\section{Deconvolution of Raman spectra}
\label{SI-fit}

To extract the integrated intensity of a spectral feature, we fit the measured Stokes-shift Raman spectra with the product of the Bose-Einstein distribution and a multi-damped Lorentz oscillator line shape,

\begin{equation}
I_{Raman}(\Omega) = \bigg( \frac{1}{e^{\frac{\hbar \Omega}{k_{B}T}    }-1} +1 \bigg) \sum_{i} \frac{c_{i} |\Omega| \Gamma_{i}^{3}}{\Omega ^{2} \Gamma_{i}^{2} +(\Omega^{2} -\omega _{i}^{2})^{2}}
\label{eq:lorentz}
\end{equation}
Where $\omega_{i}$, $c_{i}$ and $\Gamma_{i}$ are the position, intensity, and width of each peak, respectively. 
$\Omega$ is the measured frequency (Raman shift), $T$ is the temperature, $\hbar$ is the Planck constant and $k_{B}$ is the Boltzmann constant. 
The Lorentz in Equation \eqref{eq:lorentz} is a variation of the Lorentz oscillator, where $c$ is the maximum value of the peak.
Given the fitted Lorentzian parameters, the integrated intensity of the spectral feature is
\begin{equation}
    I_{int}=\left[n_{\textrm{BE}}(\omega_{i},T)+1\right]\frac{2c_{i}\left|\omega_i\right|\Gamma_{i}^2}{\sqrt{4\omega_{i}^2-\Gamma_{i}^2}}\left[\frac{\pi}{2}-\tan^{-1}\left(\frac{\Gamma_{i}^2-2\omega_{i}^2}{\Gamma_{i}\sqrt{4\omega_{i}^2-\Gamma_{i}^2}}\right)\right],
\end{equation}
with $n_{\textrm{BE}}$ standing for an approximated Bose-Einstein constant pre-factor.

\pagebreak

\section{\label{Si_PO_SI}The temperature independent polarization-orientation Raman of silicon}

\begin{figure*}[h]
\centering
\includegraphics[scale=0.6]{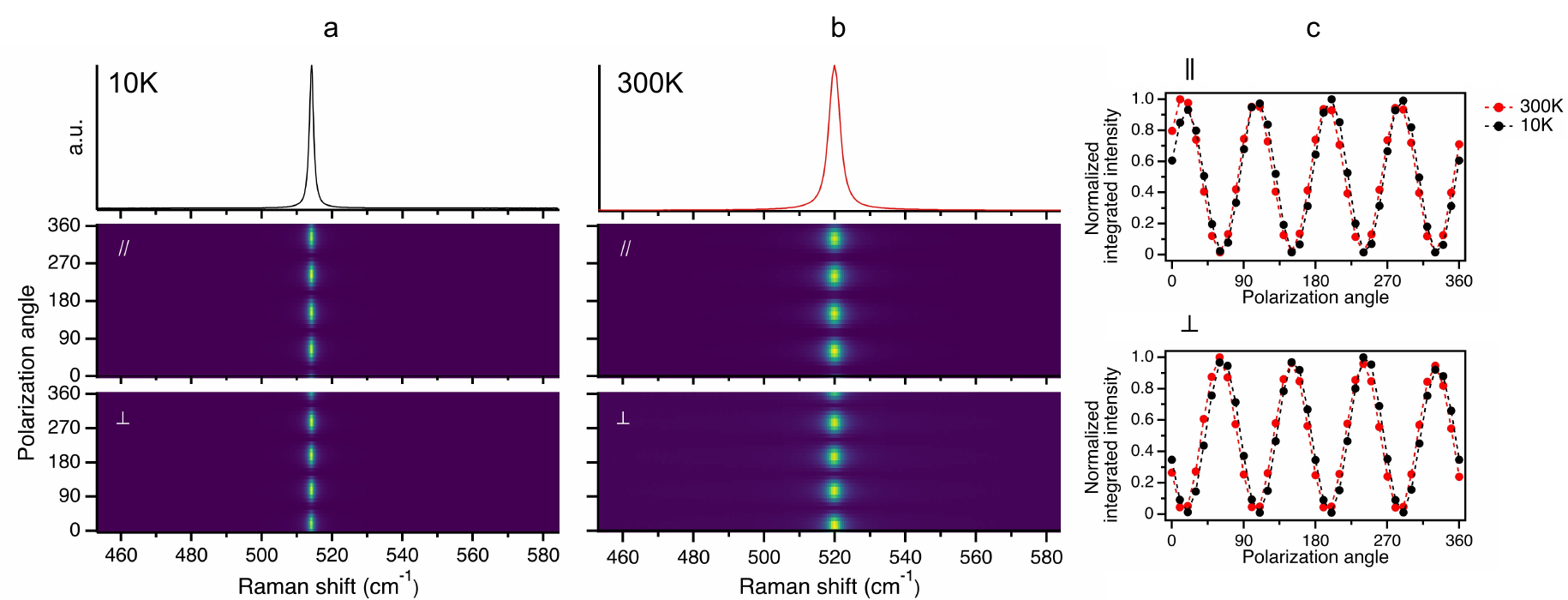}
%[width=8.5 cm]
\caption{\label{fig:Si_PO}Temperature dependence of the PO Raman of silicon. (a) and (b) presents the PO Raman measurement of silicon at 10~K and 300~K, respectively. The middle and bottom panels present the measurements in parallel and cross configurations, respectively. The top panel presents the unpolarized spectrum (sum over all angles). (c) presents the polarization dependence of the integrated intensity of the prominent Raman peak. The top and bottom panels present the results for parallel and cross configurations, respectively.}
\end{figure*}

Figure~\ref{fig:Si_PO} presents the PO Raman measurements of a (100) silicon wafer at 10~K and 300~K with the analysis of the polarization dependence of the integrated intensity of the prominent Raman peak (the TO phonon at around 520~cm$^{-1}$).
We performed the measurements and analysis similar to the organic crystals discussed in the main text using the same optical setup. 
We present these results as an example for an inorganic material that shows a temperature independent PO response, in agreement with the harmonic picture.
Furthermore, according to theory, the intensity of the prominent Raman peak should go down to zero at the minimum point~\cite{Lu2005}.
In our measurement, the peak intensity drops to about 1\% of its maximum intensity, showing our system suffers very little leakage.

\clearpage

\section{\label{chloro_PO_SI} Polarization-orientation Raman of chloroform}

\begin{figure*}[h]
\centering
\includegraphics[scale=0.9]{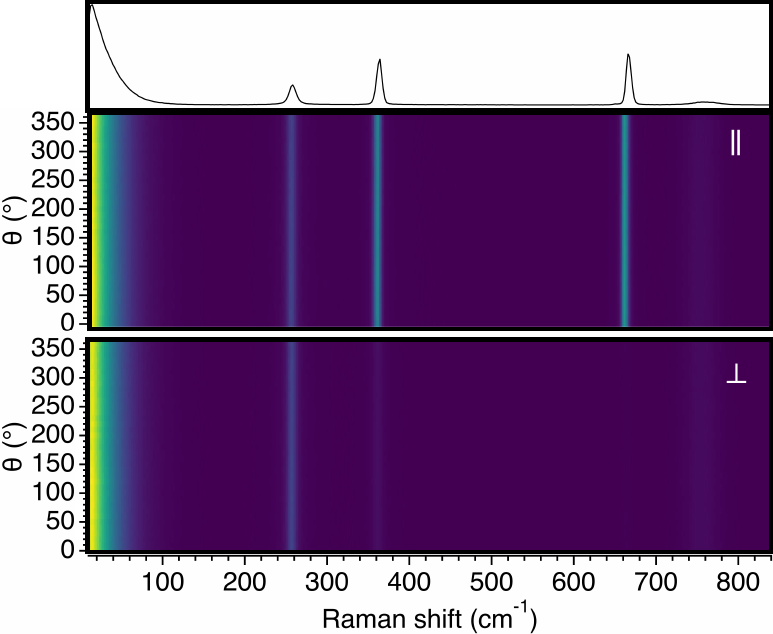}
\caption{\label{fig:Chloroform_PO}Polarization dependent measurement of liquid chloroform in parallel and cross configurations. 
The top panel shows a typical spectrum in the parallel configuration.}
\end{figure*}

Figure~\ref{fig:Chloroform_PO} presents the PO Raman measurement of liquid chloroform at room temperature.
The measurement was performed similarly to the organic crystals discussed in the main text using the same optical setup. 
Since there is no long-range order in a liquid, we expect the measurement to be polarization independent.
This is why measuring the PO response of a liquid is helpful as a test for the system response to the change in polarization angle.
Our results show that the PO response of the vibrations of chloroform is completely polarization-independent - proving our systems response is minimal.

\clearpage

\section{\label{SI_BTBT_T_dependent}Temperature dependent low-frequency Raman spectroscopy of BTBT}

\begin{figure*}[h]
\centering
\includegraphics[scale=0.85]{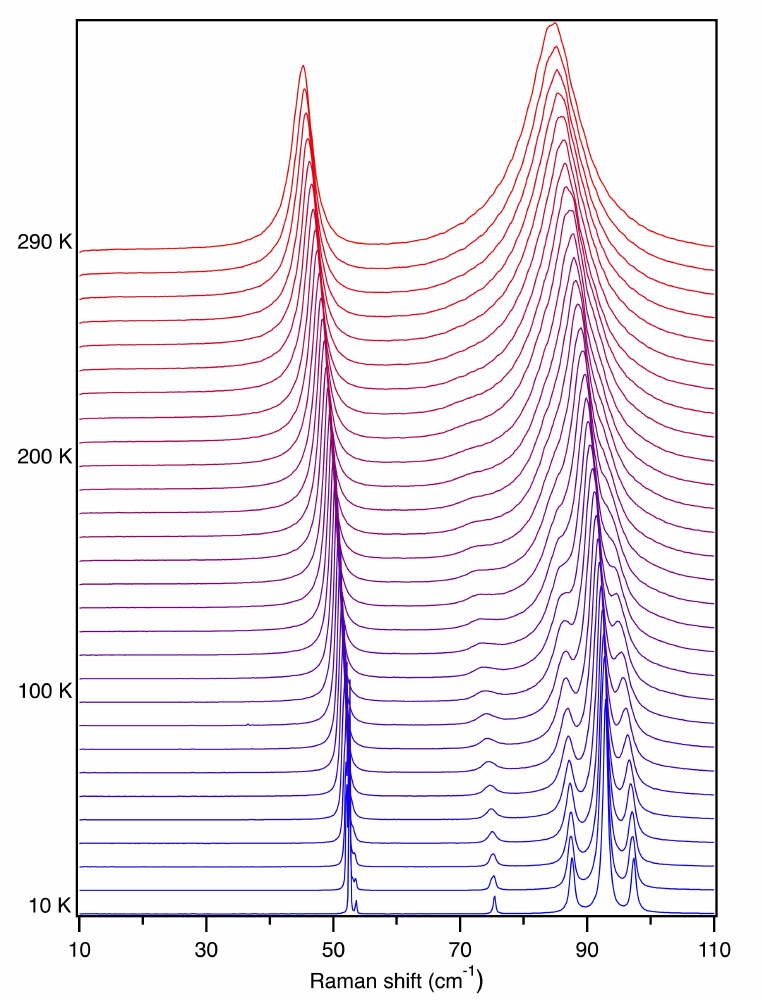}
\caption{\label{fig:T_dep_SI}Temperature dependent low-frequency Raman of BTBT. The spectra were normalized and shifted up for clarity. The temperature increment is 10~K.}
\end{figure*}

Figure~\ref{fig:T_dep_SI} shows the temperature dependent low-frequency Raman spectroscopy spectra of BTBT, from 10~K to 290~K at 10~K increments.
We see only a gradual redshift and broadening of the spectrum as temperature increases.
No abrupt changes to the Raman spectrum indicate there is no phase transition in the measured temperature range.

\begin{figure*}[ht!]
\centering
\includegraphics[scale=0.9]{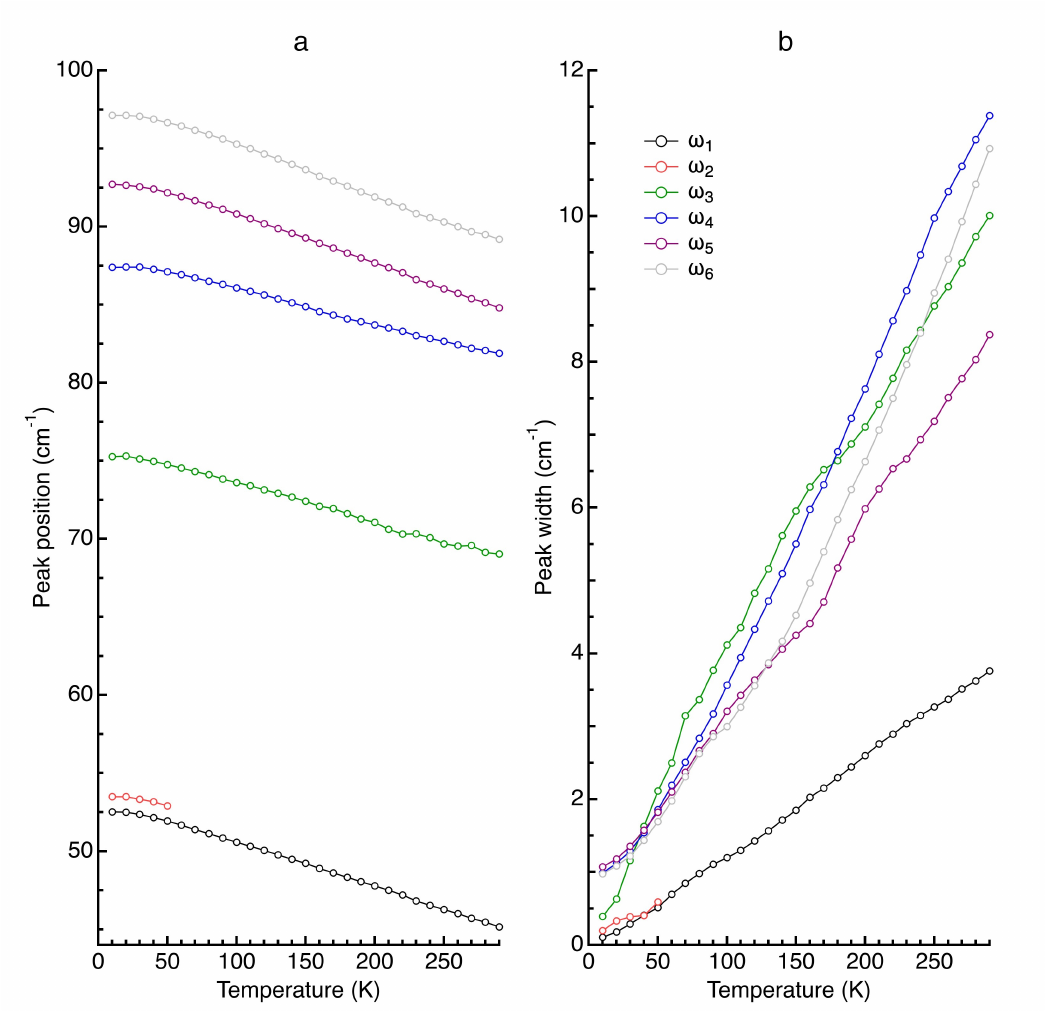}
\caption{\label{fig:w_g_vs_T}The temperature dependence of the vibrational frequencies (a) and FWHM (b) of the lattice vibrations of BTBT.}
\end{figure*}

Figure~\ref{fig:w_g_vs_T} shows the temperature dependence of the peaks position and full-width-half-maximum (FWHM) of the lattice vibrations, extracted from fitting the spectra according to Sec.~\ref{SI-fit}.
The trend with temperature of vibrational frequency and FWHM is primarily linear.

\clearpage

\section{\label{temp_PO_all_SI}Raman polarization-orientation color maps of BTBT}

\begin{figure*}[h]
\centering
\includegraphics[scale=0.65]{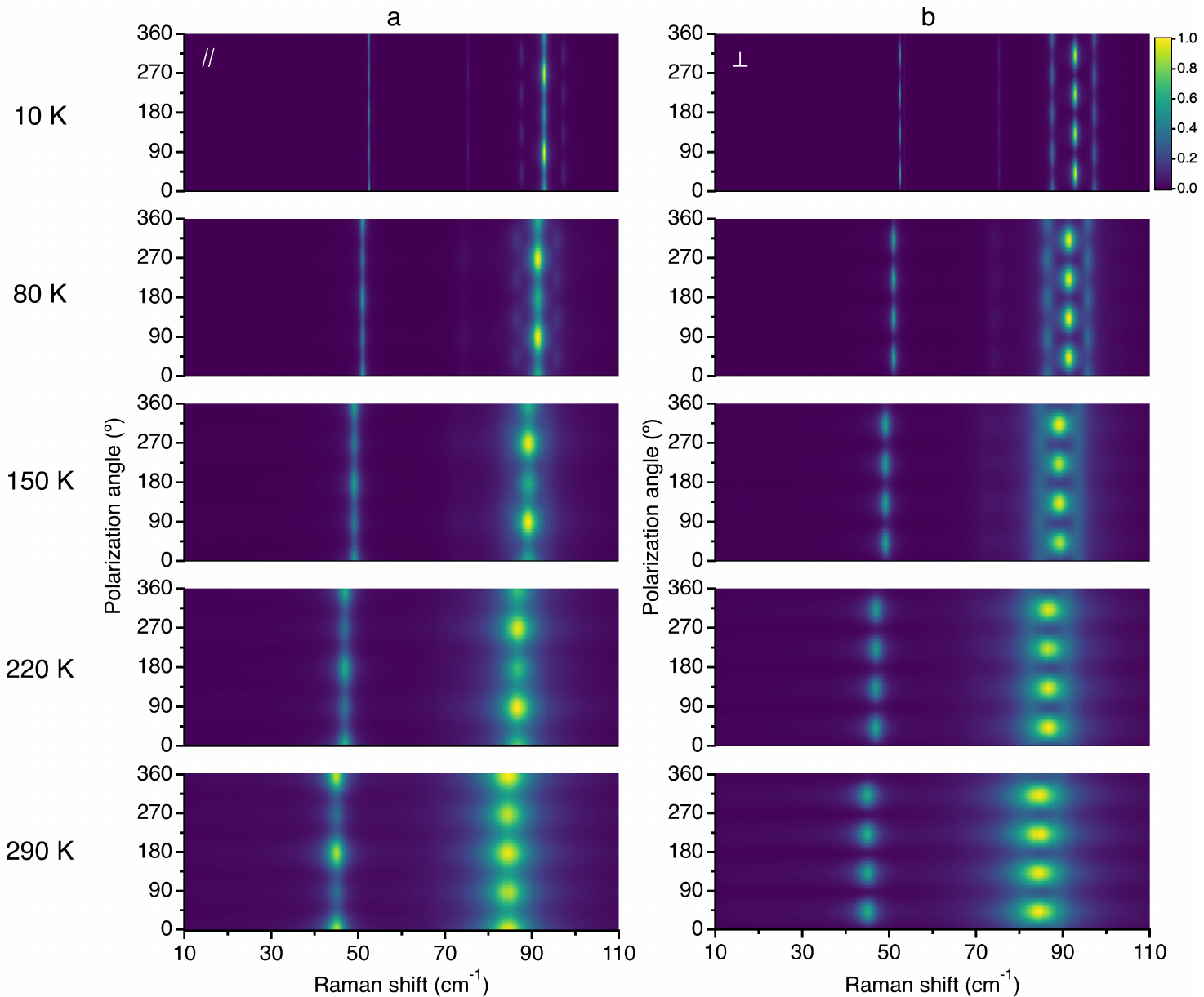}
\caption{\label{fig:Raw_PO_BTBT}Raw PO Raman of BTBT in (a) parallel and (b) cross configurations at 10~K, 80~K, 150~K, 220~K and 290~K.}
\end{figure*}

Figure~\ref{fig:Raw_PO_BTBT} presents the raw data obtained from the PO Raman measurements for both parallel and cross configurations for a single crystal of BTBT at 10, 80, 150, 200, and 290~K.

\clearpage

\section{\label{T_dep_PO_SI}Temperature evolution in the Raman polarization-orientation response of BTBT}

\begin{figure*}[h]
\centering
\includegraphics[scale=0.9]{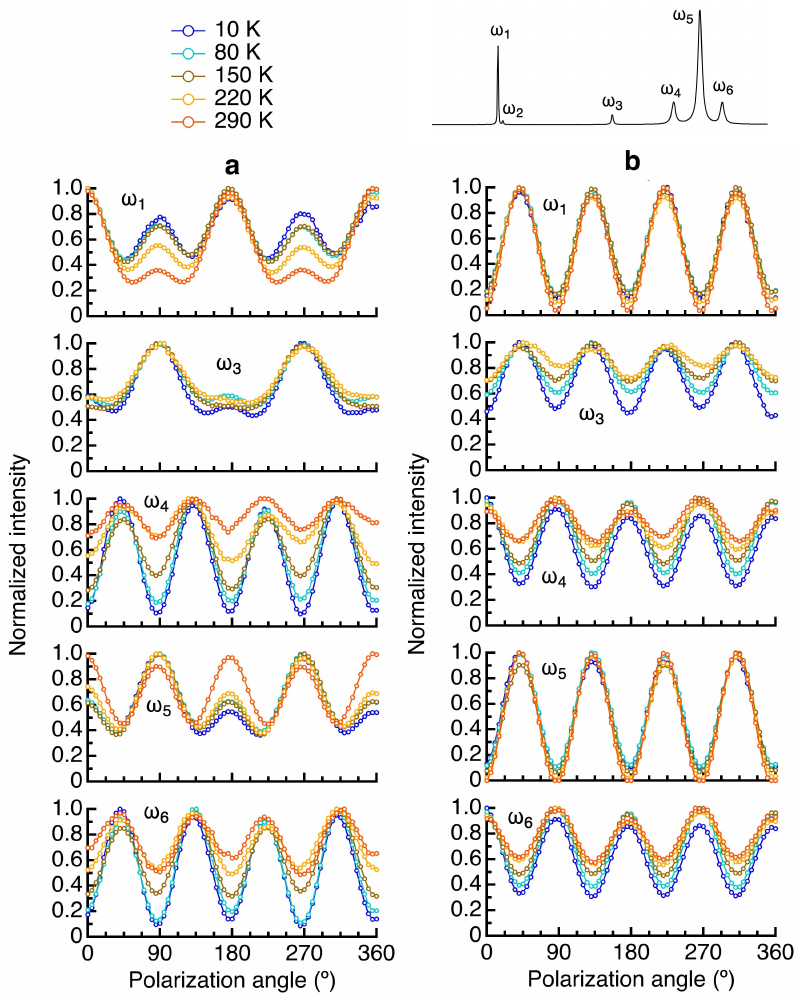}
\caption{\label{fig:PO_response}The temperature evolution of the PO dependence for the low-frequency peaks of BTBT in (\textbf{a}) parallel and (\textbf{b}) cross configurations. The intensities were normalized for each mode at each temperature.}
\end{figure*}

Figure~\ref{fig:PO_response} presents the temperature evolution of the PO pattern of each separate peak in the Raman spectrum of BTBT.
The results show the PO Raman response of all BTBT lattice vibrations evolves with temperature.

\clearpage

\section{\label{SI_BTBTPO_fit}Fitting the PO response of BTBT}

\vspace{5pt}

\subsection{Data preparation and factor group analysis for BTBT}

Section~\ref{temp_PO_all_SI} presents the contour plots of the PO Raman measurement of BTBT at 10, 80, 150 220, and 290~K in parallel and cross configurations.
The measurements were performed as described in Sec.~\ref{SI-experiment}.
To extract the polarization dependence of the integrated intensity of each peak we fit the spectra at each polarization angle to a multi-damped Lorentz oscillator (see section~\ref{SI-fit}).
The dotted lines in Figures~\ref{fig:SI_BTBT_PO_fit} and~~\ref{fig:SI_BTBT_4_rank_fit} show the integrated intensity of each damped Lorentz oscillator with respect to the excitation polarization angle.
Next, for the second-rank Raman tensor formalism we extract the form of the Raman tensors, as well as the expected number of lattice modes and their vibrational symmetry, by using factor group analysis with the relevant space group (monoclinic, $P2_{1}/c$)~\cite{Porto1981}.
For BTBT, factor group analysis predicts 6 Raman-active lattice libration modes, and indeed our measured spectra showed exactly six distinguishable Raman peaks.

\vspace{5pt}

\subsection{Fitting the PO with a second-rank tensor}
\label{2ndRank_fit}

The symmetry allowed Raman tensors for $3\text{A}_{g}$ and $3\text{B}_{g}$ are\cite{Bilbao2006}:
\begin{equation}
 \mathcal{R}_{\mathrm{A_{g}}}=\begin{pmatrix} a&0&e\\0&b&0\\e&0&c \end{pmatrix},~~~~~~~~~~\mathcal{R}_{\mathrm{B_{g}}}=\begin{pmatrix} 0&d&0\\d&0&f\\0&f&0 \end{pmatrix}.
\label{eq:Raman_tensor}
\end{equation}

We perform a global fit to the integrated intensity in both parallel and cross configurations using the second-rank Raman tensor formalism described by Eq.~$(1)$ in the main text (see Ref.~\cite{Asher2020} for more details regarding the fitting process).

\begin{figure*}[h]
\centering
\includegraphics[scale=0.55]{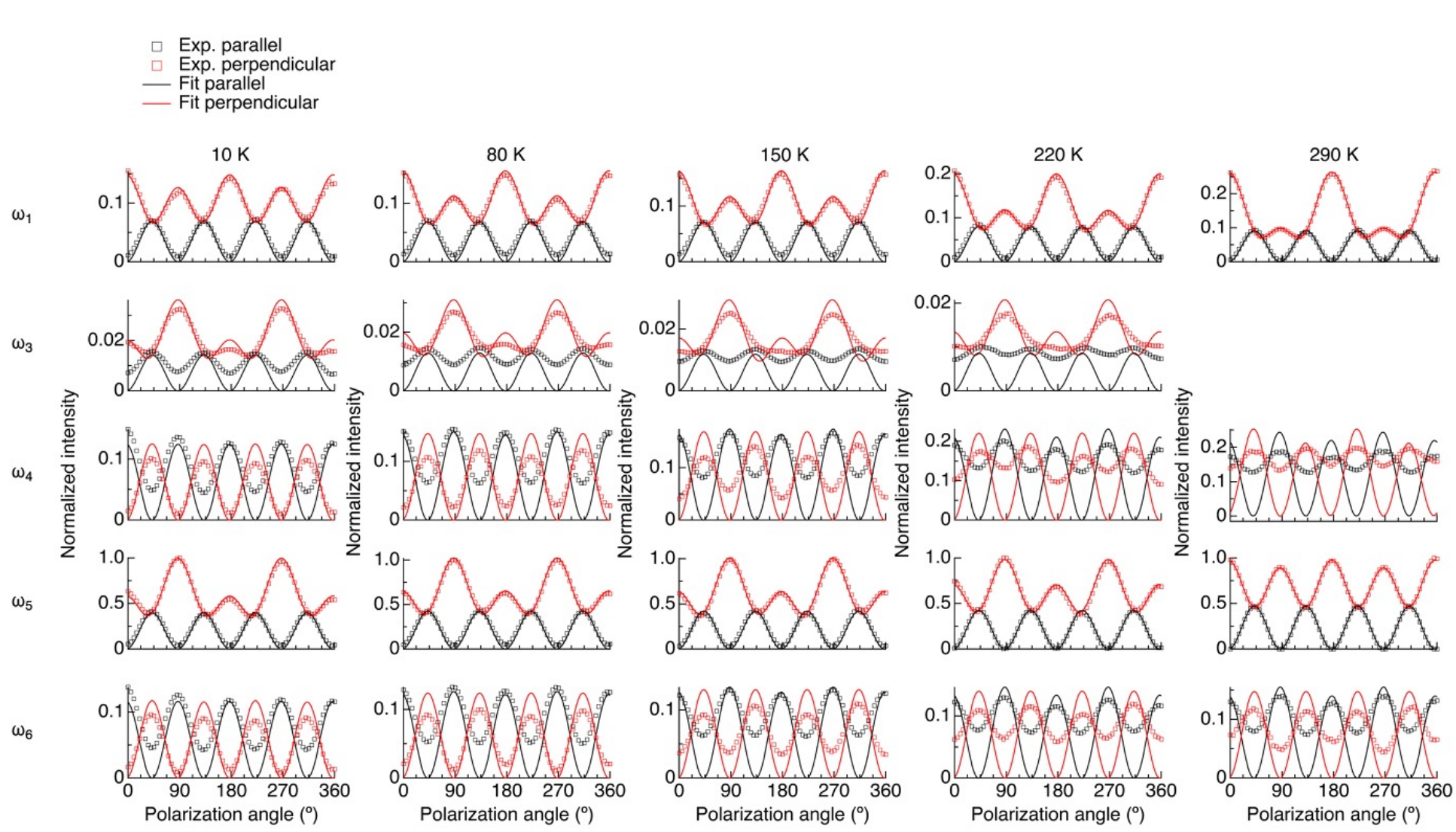}
\caption{\label{fig:SI_BTBT_PO_fit}Fitting results for the PO Raman response of BTBT at different temperatures, using a general second-rank Raman tensor.}
\end{figure*}

Figure \ref{fig:SI_BTBT_PO_fit} shows the results for this fitting procedure at all measured temperatures.
The 10~K fits show that $\omega_{1}$, $\omega_{3}$, and $\omega_{5}$ are A$_{g}$ modes, while $\omega_{4}$ and $\omega_{6}$ are B$_{g}$ modes.
The intensity of $\omega_{2}$ was too weak to extract its polarization dependence reliably.
The fit gets worse for $\omega_{3}$, $\omega_{4}$ and $\omega_{6}$ as temperature increases, while for $\omega_{1}$ and $\omega_{5}$ the fit is successful in all temperatures.

To further challenge the necessity of a fourth-rank tensor, we attempted fitting a symmetry-relaxed second-rank tensor:
\begin{equation}
    \tilde{\mathcal{R}}=\left(\begin{array}{ccc}
    a & d_1 & e_1\\
    d_2 & be^{i\delta} & f_1\\
    e_2 & f_2 & c
\end{array}\right),
\end{equation}
where we introduce a relative phase $\delta$ to account for possible birefringence effects~\cite{Kranert2016}, and allow asymmetric components (effectively relaxing time-symmetry)~\cite{Yu2010}.
Adapted to our back-scattering geometry, $\tilde{\mathcal{R}}$ has five independent parameters $(a,b,d_1,d_2,\delta)$.
As Fig.~\ref{fig:SI_BTBT_PO_fit} demonstrates, this generalized second-rank tensor form was still unable to fit the observed PO pattern of $\omega_4$.

\vspace{5pt}

\subsection{Fitting the PO with a fourth-rank tensor}
\label{4rnk_fit}

\begin{figure*}[h]
\centering
\includegraphics[scale=0.7]{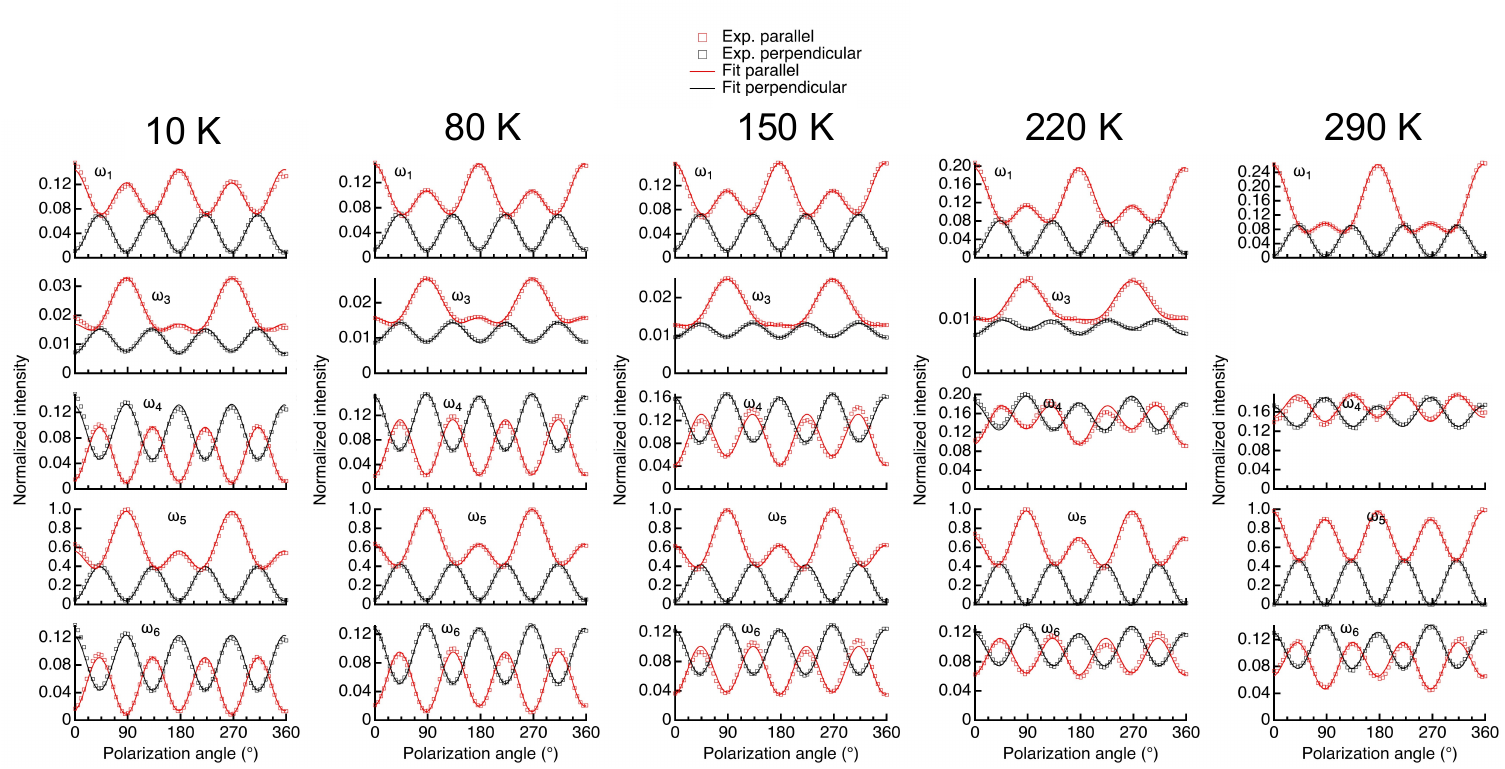}
\caption{\label{fig:SI_BTBT_4_rank_fit}Fitting results for the PO Raman response of BTBT at different temperatures, showing the successful fit obtained by using the fourth-rank formalism.}
\end{figure*}

Applying the symmetry constraints of the relevant space group $(P2_{1}/c)$ along with the intrinsic left and right minor symmetries, and retaining only the components relevant for our back-scattering geometry, the effective fourth-rank tensor used for fitting the PO patterns was:
\begin{equation}
\label{eq:4rnk_tensor}
I_{\mu\nu\xi\rho}=\left(\begin{array}{ccc}
\left(\begin{array}{ccc}
a & 0 & 0\\
0 & b & 0\\
0 & 0 & 0
\end{array}\right) & \left(\begin{array}{ccc}
0 & d & 0\\
d & 0 & 0\\
0 & 0 & 0
\end{array}\right) & \left(\begin{array}{ccc}
0 & 0 & 0\\
0 & 0 & 0\\
0 & 0 & 0
\end{array}\right)\\
\left(\begin{array}{ccc}
0 & d & 0\\
d & 0 & 0\\
0 & 0 & 0
\end{array}\right) & \left(\begin{array}{ccc}
b & 0 & 0\\
0 & c & 0\\
0 & 0 & 0
\end{array}\right) & \left(\begin{array}{ccc}
0 & 0 & 0\\
0 & 0 & 0\\
0 & 0 & 0
\end{array}\right)\\
\left(\begin{array}{ccc}
0 & 0 & 0\\
0 & 0 & 0\\
0 & 0 & 0
\end{array}\right) & \left(\begin{array}{ccc}
0 & 0 & 0\\
0 & 0 & 0\\
0 & 0 & 0
\end{array}\right) & \left(\begin{array}{ccc}
0 & 0 & 0\\
0 & 0 & 0\\
0 & 0 & 0
\end{array}\right)
\end{array}\right)
\end{equation}
Figure~\ref{fig:SI_BTBT_4_rank_fit} shows the fit results for the same PO Raman data set by using the fourth-rank Raman tensor formalism.
Here we see that the fit is excellent at all temperatures.
These results are discussed in the main text.

\pagebreak

\subsection{Global fit with the two-mode model}

The full PO data for spectral features $\omega_4$ and $\omega_5$ at 10~K was globally fitted using the two-mode model discussed in the main text, with birefringence effects incorporated as in the harmonic case of Sec.~S9.2, introducing the additional parameter $\delta$. 
%introducing an additional parameter $\delta$ into the susceptibility tensors.

In practice, this means \emph{simultaneously} fitting the spectral range 85-95 $cm^{-1}$ for all 144 spectra (72 polarization angles for each polarization configuration) to:
\begin{equation}
\label{eq:fit}
    \sigma(\Omega) \propto \sum_{\mu\nu\xi\rho}n_{\mu}n_{\xi}I_{\mu\nu\xi\rho}(\Omega)E_{\nu}E_{\rho},
\end{equation}
with
\begin{equation}
     I_{\mu\nu\xi\rho}(\Omega)= 
     \left[n_{BE}(\Omega,T)+1\right]
     \left[{\chi^{*}}^{\mu\nu}_{\mathrm{A_g}}\chi^{\xi\rho}_{\mathrm{A_g}} J_{\mathrm{A_gA_g}} + 
     {\chi^{*}}^{\mu\nu}_{\mathrm{B_g}}\chi^{\xi\rho}_{\mathrm{B_g}} J_{\mathrm{B_gB_g}}\right],
\end{equation}
and
\begin{equation}
 \chi_{\mathrm{A_{g}}}=\begin{pmatrix} a&0&e\\0&be^{i\delta}&0\\e&0&c \end{pmatrix},~~~~~~~~~~\chi_{\mathrm{B_{g}}}=\begin{pmatrix} 0&de^{i\delta}&0\\de^{i\delta}&0&fe^{i\delta}\\0&fe^{i\delta}&0 \end{pmatrix}.
\label{eq:Raman_tensor}
\end{equation}
Applying our back-scattering geometry to Eq.~\eqref{eq:fit} gives the following final fitted expressions for parallel $(\parallel)$ and cross $(\perp)$ configurations:
\begin{multline}
      \sigma_{\parallel} \propto J_{\mathrm{AgAg}} a^2 \cos^{4}(\Delta\theta) + 2 J_{\mathrm{AgAg}} a b \cos(2\delta) \cos^{2}(\Delta\theta) \sin{^2}(\Delta\theta)^2 +\\
      4 J_{\mathrm{BgBg}} d^2 \cos^{2}(\Delta\theta) \sin^{2}(\Delta\theta) + J_{\mathrm{AgAg}} b^2 \sin^{4}(\Delta\theta), 
\end{multline}
\begin{multline}
      \sigma_{\perp} \propto J_{\mathrm{AgAg}} a^2 \cos^{2}(\Delta\theta) \sin{^2}(\Delta\theta)^2 + J_{\mathrm{AgAg}} b^2 \cos^{2}(\Delta\theta) \sin{^2}(\Delta\theta)^2 -\\ 
      2 J_{\mathrm{BgBg}} d^2 \cos^{2}(\Delta\theta) \sin^{2}(\Delta\theta) - 2 J_{\mathrm{AgAg}} a b \cos(2\delta) \cos^{2}(\Delta\theta) \sin{^2}(\Delta\theta)^2 +\\ 
      J_{\mathrm{BgBg}} d^2 \cos^{4}(\Delta\theta) + J_{\mathrm{BgBg}} d^2 \sin^{4}(\Delta\theta),  
\end{multline}
where $\Delta\theta=\theta-\theta_0$ is determined by the incident polarization $\theta$, controlled in experiment, and $\theta_0$ is a global angle determined by the crystal orientation.
The spectral frequency dependency is contained within the spectral functions $J_{\lambda\lambda'}(\Omega)$ and the $ \left[n_{BE}(\Omega,T)+1\right]$ factor. 

The full derivation of $J_{\lambda\lambda'}$ for the two-mode model is given in the main text, equations $(6),(7),(9)$.
This yielded the final function used in the global fit (using the shorthand $\mathrm{A_g}\equiv 1$ and $\mathrm{B_g}\equiv 2$):
\begin{equation}
    J_{11}(\Omega)=
    \frac{
    \Gamma_{2}\left(-\gamma^{*}\gamma-\Gamma_{1}\Gamma_{2} + \left(\frac{\Omega^2-\omega_{1}^2}{2\omega_{1}}\right)\left(\frac{\Omega^2-\omega_{2}^2}{2\omega_{2}}\right)\right) - 
    \left(\frac{\Omega^2-\omega_{2}^2}{2\omega_{2}}\right) \left(\Gamma_{1}\left(\frac{\Omega^2-\omega_{2}^2}{2\omega_{2}}\right) + \Gamma_{2}\left(\frac{\Omega^2-\omega_{1}^2}{2\omega_{1}}\right) \right)
    } 
    {\left(-\gamma^{*}\gamma - \Gamma_{1}\Gamma_{2} + \left(\frac{\Omega^2-\omega_{1}^2}{2\omega_{1}}\right) \left(\frac{\Omega^2-\omega_{2}^2}{2\omega_{2}}\right) \right)^2 +
    \left( \Gamma_{1}\left(\frac{\Omega^2-\omega_{2}^2}{2\omega_{2}}\right) + \Gamma_{2}\left(\frac{\Omega^2-\omega_{1}^2}{2\omega_{1}}\right) \right)^2
    },
\end{equation}
with $J_{\mathrm{B_gB_g}}$ being the same only with swapped irrep indices $\left(\mathrm{A_g}	\leftrightarrow\mathrm{B_g}\right)$.  

The full list of fit parameters is therefore:
\begin{enumerate}
    \item $\omega_i$ - the peak resonance
    \item $\Gamma_i$ - the peak width
    \item $\theta_0$ - global incident polarization off-set
\end{enumerate}    
The three parameters mentioned above are identical to those used to extract the integrated intensity in the harmonic analysis. 
They govern the peak shape, but not the PO pattern. 
The parameters that create the PO response in our two-mode model are: 
\begin{enumerate}[resume]
    \item $a$ - susceptibility derivative for $\mathrm{A_g}$ normal mode
    \item $b$ - susceptibility derivative for $\mathrm{A_g}$ normal mode
    \item $d$ - susceptibility derivative for $\mathrm{B_g}$ normal mode
    \item $\delta$ - phase retardance due to birefringence
    \item $\gamma$ - off-diagonal self-energy component
\end{enumerate}
which gives a total of five parameters for two spectral peaks, \emph{the same} as in the generalized second-rank formalism that failed to account for the PO response of the \emph{single} $\omega_4$ peak.

\clearpage

\subsection{Positive semi-definite fourth-rank tensors}

Unlike the harmonic expression (Eq.~(1) in main text), using the general fourth-rank tensor of Eq.~\eqref{eq:fit} can yield non-positive values.  
This poses no problem when used for fitting experimental data, but since scattering cross-sections can not be negative, this technically places further constraints on the general tensor form.
Because our fourth-rank tensor has the same symmetry of the elastic stiffness tensor, we can apply the positive semi-definiteness constraint by adopting Voigt notation, where the tensor is re-written as a 6X6 matrix~\cite{nye1985,Slezak2020}.

We begin by defining the extended vectors $\tilde{\vec{n}}$ and $\tilde{\vec{E}}$:
\begin{equation}
    \tilde{\boldsymbol{n}}=\begin{pmatrix}n_{x}^{2}\\
    n_{y}^{2}\\
    n_{z}^{2}\\
    n_{y}n_{z}\\
    n_{x}n_{z}\\
    n_{x}n_{y}
    \end{pmatrix};\qquad\qquad\qquad\qquad\qquad\tilde{\boldsymbol{E}}=\begin{pmatrix}E_{x}^{2}\\
    E_{y}^{2}\\
    E_{z}^{2}\\
    E_{y}E_{z}\\
    E_{x}E_{z}\\
    E_{x}E_{y}
    \end{pmatrix}.
\end{equation}
If we write the following 6X6 matrix in terms of the fourth-rank tensor terms:
\begin{equation}
\label{eq:voigtmat}
    \tilde{I}=\begin{pmatrix}I_{1111} & I_{1212} & I_{1313} & 2I_{1213} & 2I_{1113} & 2I_{1112}\\
    I_{1212} & I_{2222} & I_{2323} & 2I_{2223} & 2I_{1223} & 2I_{1222}\\
    I_{1313} & I_{2323} & I_{3333} & 2I_{2333} & 2I_{3313} & 2I_{1323}\\
    2I_{1213} & 2I_{2223} & 2I_{2333} & 2\left(I_{2233}+I_{2323}\right) & 2\left(I_{1233}+I_{1323}\right) & 2\left(I_{1223}+I_{1322}\right)\\
    2I_{1113} & 2I_{1223} & 2I_{1333} & 2\left(I_{1233}+I_{1323}\right) & 2\left(I_{1133}+I_{1313}\right) & 2\left(I_{1123}+I_{1213}\right)\\
    2I_{1112} & 2I_{1222} & 2I_{1323} & 2\left(I_{1223}+I_{1322}\right) & 2\left(I_{1123}+I_{1213}\right) & 2\left(I_{1122}+I_{1212}\right)
    \end{pmatrix},
\end{equation}      
then the expression 
\begin{equation}
\label{eq:voigtform}
    \tilde{I}=\tilde{\vec{n}}\cdot\tilde{I}\cdot\tilde{\vec{E}}
    =\sum_{\alpha\beta}\tilde{I}_{\alpha\beta}n_{\alpha}E_{\beta}
\end{equation}
is equivalent to the cross-section defined by Eq.~\eqref{eq:fit}.

It is now straightforward to guarantee the positive semi-definiteness of Eq.~\eqref{eq:voigtform} by demanding all eigenvalues of the symmetric matrix $\tilde{I}$ to be non-negative~\cite{VandenBos2007}.
This will generally yield intractable expressions. 
Even after applying the spatial symmetry constraints for BTBT ($C_{2h}$ point group) the scattering tensor still has $14$ independent variables with the Voigt-like matrix:
\begin{equation}
    \tilde{I}_{C_{2h}}=
    \begin{pmatrix}I_{1111} & I_{1212} & I_{1313} & 0 & 2I_{1113} & 0\\
    I_{1212} & I_{2222} & I_{2323} & 0 & 2I_{1223} & 0\\
    I_{1313} & I_{2323} & I_{3333} & 0 & 2I_{1333} & 0\\
    0 & 0 & 0 & 2\left(I_{2233}+I_{2323}\right) & 0 & 2\left(I_{1223}+I_{1322}\right)\\
    2I_{1113} & 2I_{1223} & 2I_{1333} & 0 & 2\left(I_{1133}+I_{1313}\right) & 0\\
    0 & 0 & 0 & 2\left(I_{1223}+I_{1322}\right) & 0 & 2\left(I_{1122}+I_{1212}\right)
    \end{pmatrix}
\end{equation}
which has complicated eigenvalues.

In the high-symmetry case of the $O_{h}$ point group, however, the scattering tensor has only three independent variables and the Voigt-like matrix is
\begin{equation}
    \tilde{I}_{O_h}=
    \begin{pmatrix}a & b & b & 0 & 0 & 0\\
    b & a & b & 0 & 0 & 0\\
    b & b & a & 0 & 0 & 0\\
    0 & 0 & 0 & 2\left(b+c\right) & 0 & 0\\
    0 & 0 & 0 & 0 & 2\left(b+c\right) & 0\\
    0 & 0 & 0 & 0 & 0 & 2\left(b+c\right)
    \end{pmatrix},
\end{equation}
which has only three different eigenvalues, such that the requirement for positive semi-definiteness becomes
\begin{equation}
    a-b\geq 0; \qquad b+c\geq 0; \qquad a+2b \geq 0.
\end{equation}

\clearpage

\section{\label{Ant_Pent_SI}Raman polarization-orientation measurements of anthracene and pentacene}

\begin{figure*}[ht!]
\centering
\includegraphics[width=\linewidth]{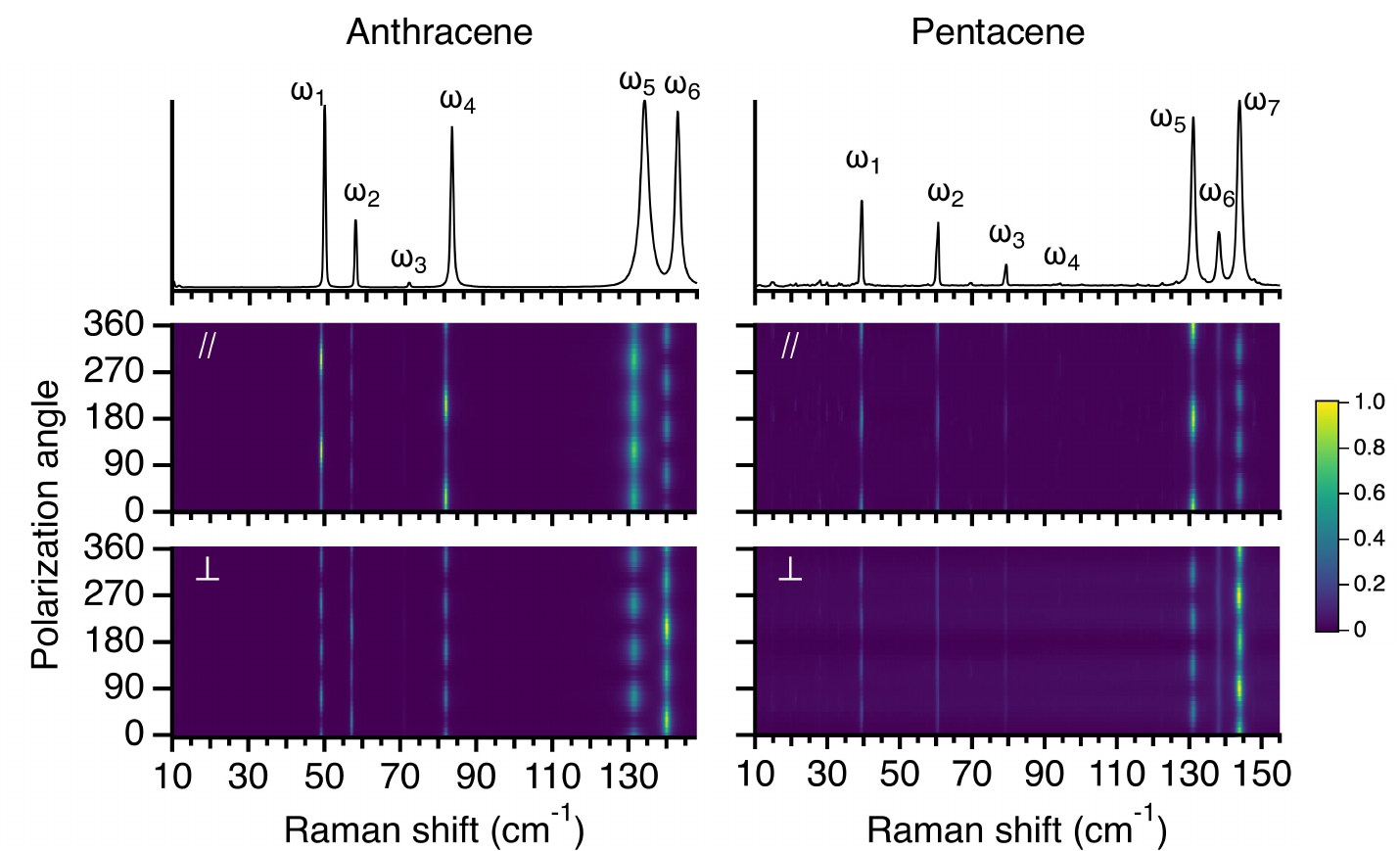}
\caption{\label{fig:ant_pent_raw}Temperature dependence of the PO Raman of anthracene (left) and pentacene (right). Top row shows the unpolarized spectra. Middle and bottom rows present the PO for parallel and cross configurations, respectively.}
\end{figure*}

The full PO Raman spectra of anthracene and pentacene crystals were measured in a similar fashion to that described in Sec.~\ref{SI-experiment} (see full measurement and irrep assignment details in \cite{Asher2020}).
The PO Raman spectra at 10~K for both anthracene and pentacene are presented in Fig.~\ref{fig:ant_pent_raw}.

\begin{figure*}[ht!]
\centering
\includegraphics[width=\linewidth]{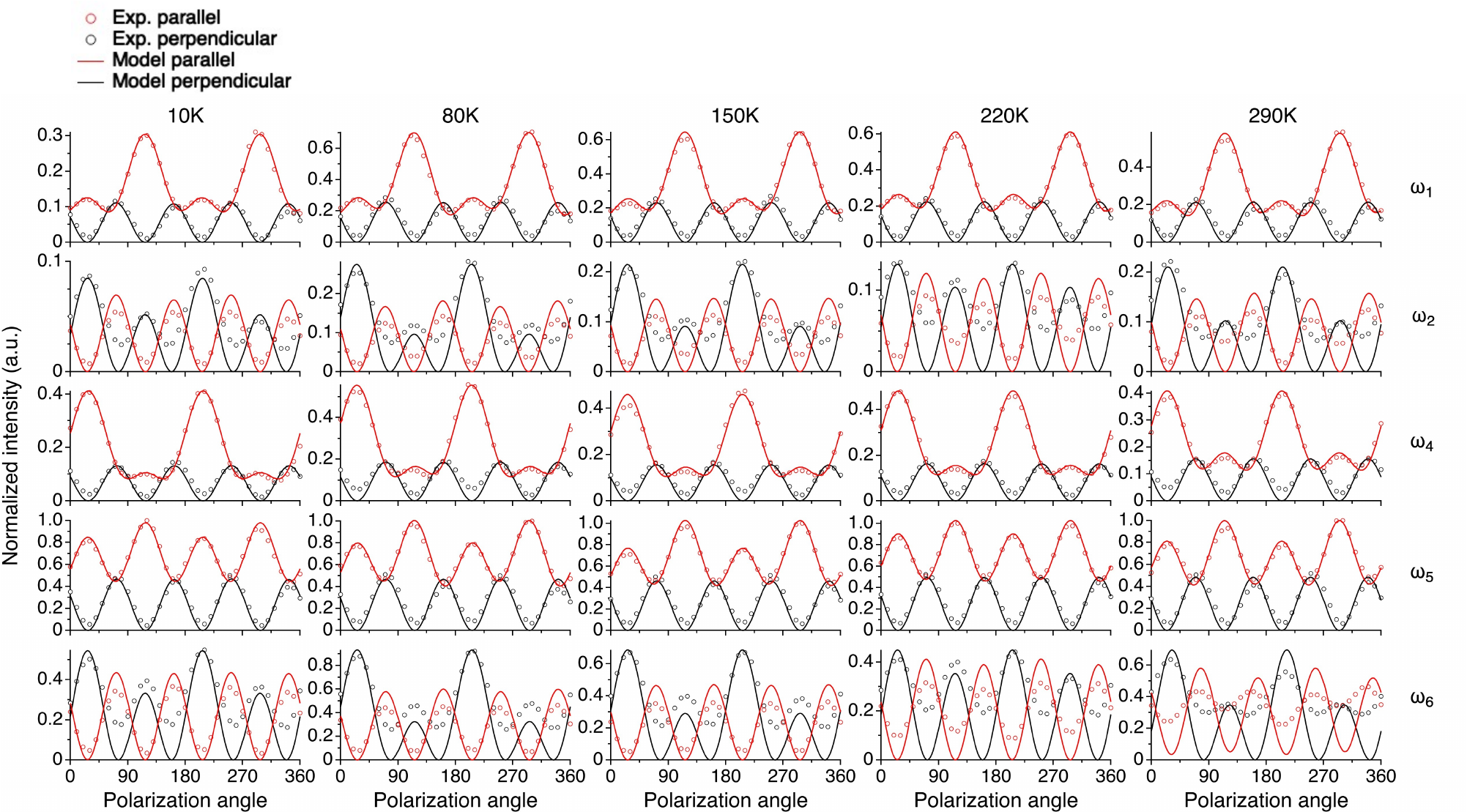}
\caption{\label{fig:ant_2rnk}Fitting results for the PO Raman response of anthracene at different temperatures, using a general second-rank Raman tensor. The fit for peaks $\omega_2$, $\omega_6$ is unsatisfactory.}
\end{figure*}

\begin{figure*}[ht!]
\centering
\includegraphics[width=\linewidth]{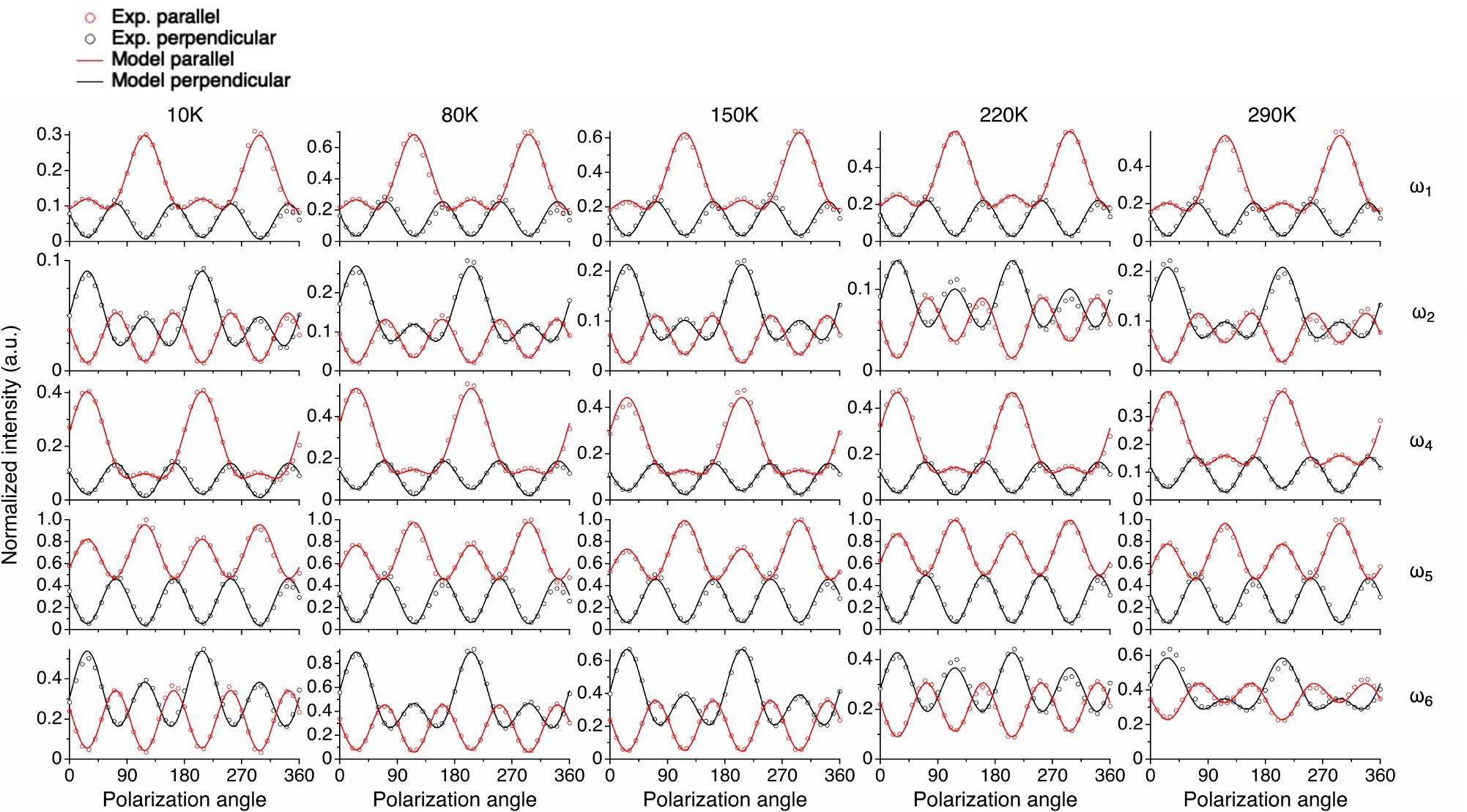}
\caption{\label{fig:ant_4rnk}Fitting results for the PO Raman response of anthracene at different temperatures, showing the successful fit obtained by using the fourth-rank formalism.}
\end{figure*}

\begin{figure*}[ht!]
\centering
\includegraphics[width=0.9\linewidth]{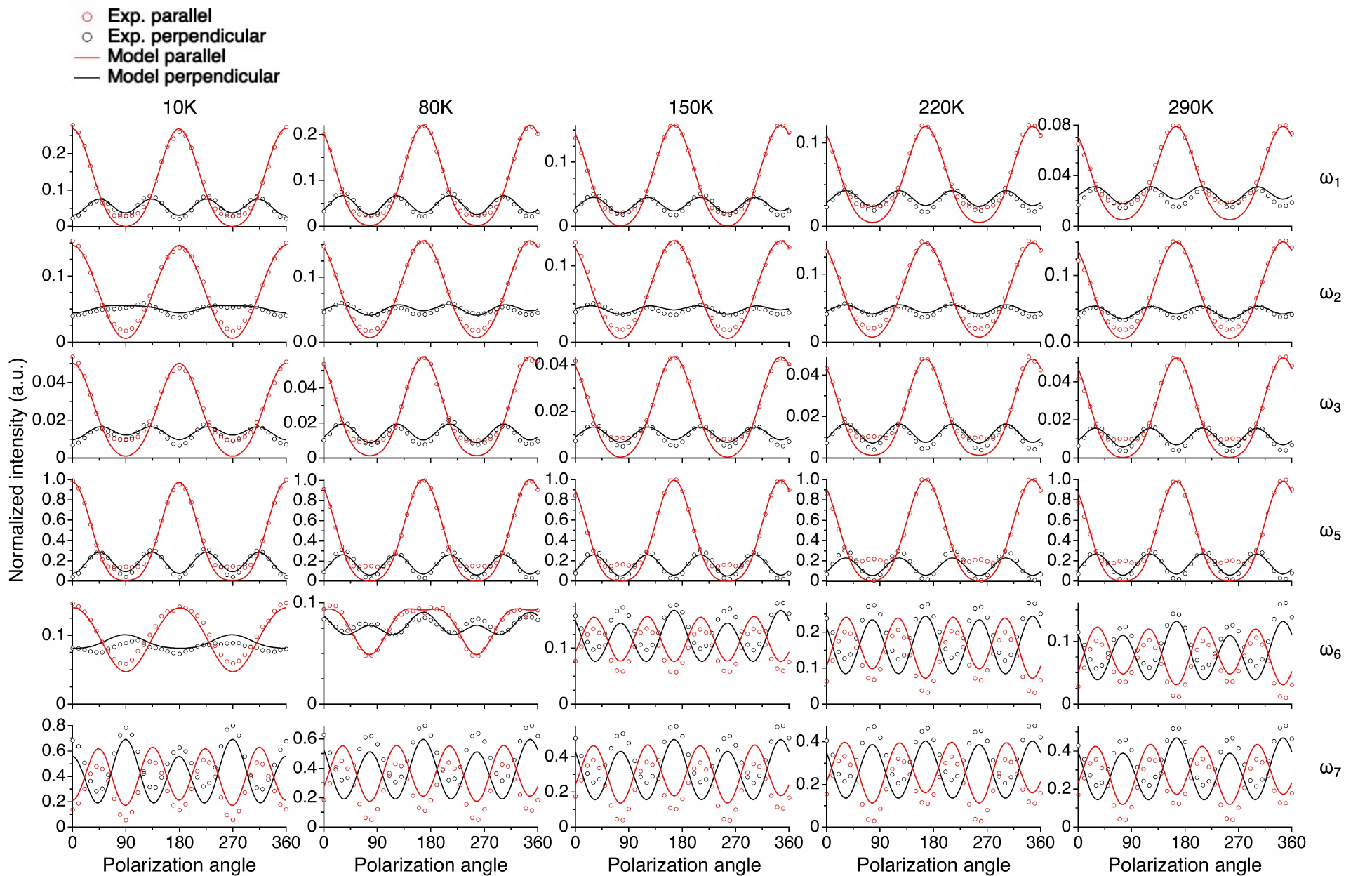}
\caption{\label{fig:pent_2rnk}Fitting results for the PO Raman response of pentacene at different temperatures, using a general second-rank Raman tensor. The fit for peaks $\omega_5$, $\omega_6$, $\omega_7$ is unsatisfactory.}
\end{figure*}

\begin{figure*}[ht!]
\centering
\includegraphics[width=0.9\linewidth]{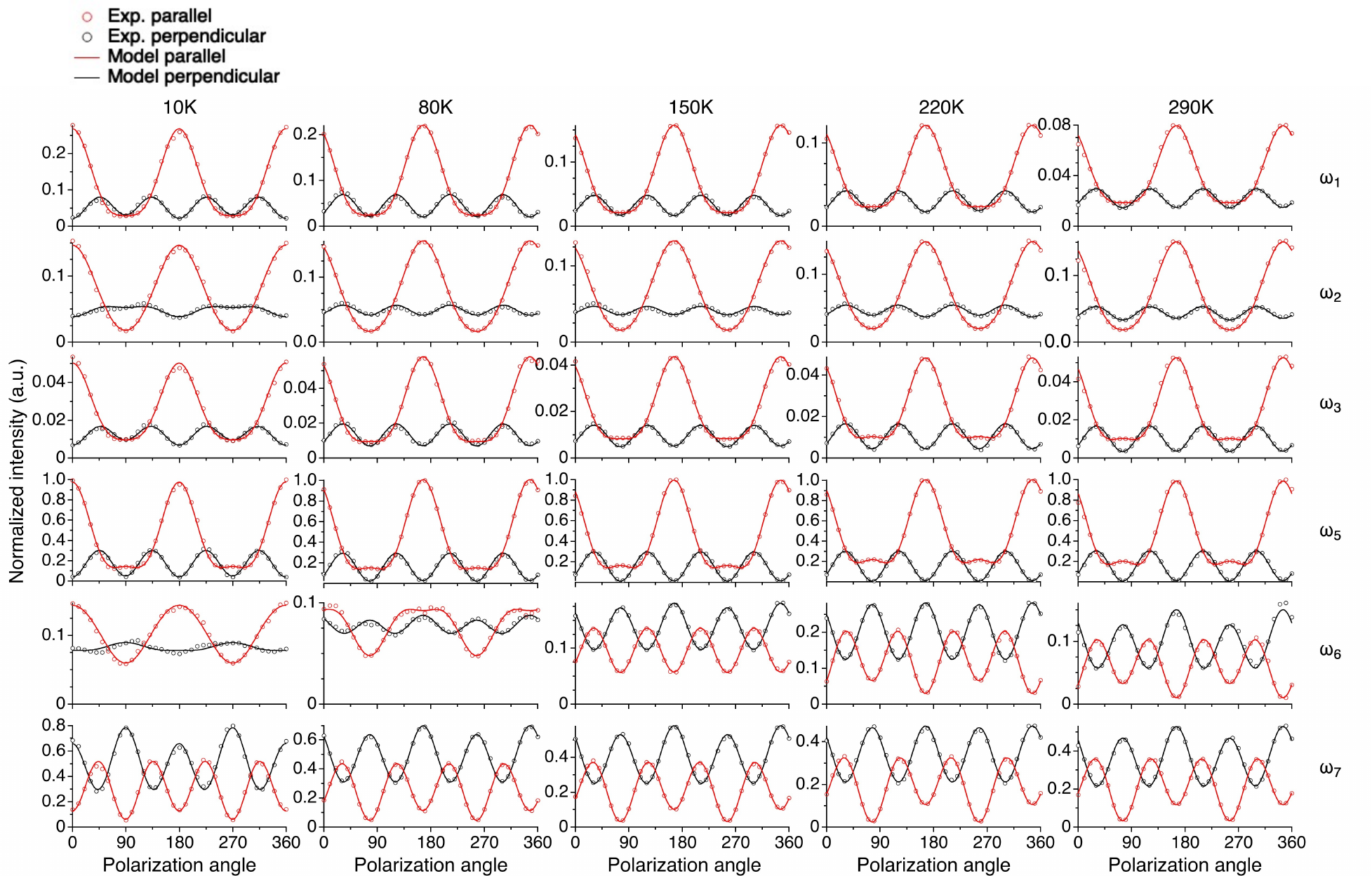}
\caption{\label{fig:pent_4rnk}Fitting results for the PO Raman response of pentacene at different temperatures, showing the successful fit obtained by using the fourth-rank formalism.}
\end{figure*}

A procedure similar to that used in BTBT (Sec.~\ref{SI-fit}) was applied to extract the integrated intensity of each peak, and fit the PO pattern to both a second (Sec.~\ref{2ndRank_fit}) and fourth-rank (Sec.~\ref{4rnk_fit}) formalism (all three crystals share the same $C_{2h}$ point-group and tensor forms).
The results are presented in figures \ref{fig:ant_2rnk} (second rank), \ref{fig:ant_4rnk} (fourth-rank) for anthracene, and figures \ref{fig:pent_2rnk} (second-rank), \ref{fig:pent_4rnk} (fourth-rank) for pentacene.
In both cases the second-rank formalism fails to capture the PO of all peaks (specifically, $\omega_2$, $\omega_6$ for anthracene, and $\omega_6$, $\omega_6$, $\omega_7$ for pentacene), whereas the fourth-rank formalism consistently yields a perfect fit.   

\clearpage

\section{Theory of inelastic light scattering from crystals at finite temperatures}
\label{Raman_Theory}
\vspace{5pt}
\subsection{Vibrational excitations of an anharmonic crystal}

In this section we outline the procedure to determine the Raman spectrum of an anharmonic crystal. 
We will begin by sketching the procedure to obtain the phonon-phonon correlation function in the presence of three-phonon anharmonicity via the equation of motion approach.

If we express the atomic displacements in terms of creation and annihilation operators we can identify the reciprocal space coefficients needed for the perturbative expansion, repeated here for clarity and consistency of notation. 
Start with the standard creation and annihilation operators for a phonon with momentum $\vec{q}$ and polarization $s$ (when there is no loss of clarity we use the compound index $\lambda$ to denote $\vec{q}s$):
\begin{align}
	a_{\vec{q}s} = & \frac{1}{\sqrt{2N\hbar}}
	\sum_{i\alpha} \epsilon_{\vec{q}s}^{i\alpha}
	\left( \sqrt{m_i \omega_{\vec{q}s}} u_{i\alpha}-i \frac{p_{i\alpha}}{ \sqrt{ m_i \omega_{\vec{q}s}} } \right) 
	e^{-i\vec{q}\cdot\vec{r}_i} 
\\
	a_{\vec{q}s}^\dagger = & \frac{1}{\sqrt{2N\hbar}}
	\sum_{i\alpha} \epsilon_{\vec{q}s}^{i\alpha\dagger} 
	\left( \sqrt{m_i \omega_{\vec{q}s}} u_{i\alpha} + i \frac{p_{i\alpha}}{ \sqrt{ m_i \omega_{\vec{q}s}} } \right) 
	e^{i\vec{q}\cdot\vec{r}_i}
\\
    A_{\vec{q}s} = a_{\vec{q}s} + a_{\bar{\vec{q}}s}^\dagger = &
\frac{2}{\sqrt{2N\hbar}}
	\sum_{i\alpha} \epsilon_{\vec{q}s}^{i\alpha}
	\sqrt{m_i \omega_{\vec{q}s}} u_{i\alpha}
	e^{-i\vec{q}\cdot\vec{r}_i} 
\\
    B_{\vec{q}s} = a_{\vec{q}s} - a_{\bar{\vec{q}}s}^\dagger = &
-\frac{2i}{\sqrt{2N\hbar}}
	\sum_{i\alpha} \epsilon_{\vec{q}s}^{i\alpha}
    \frac{p_{i\alpha}}{ \sqrt{ m_i \omega_{\vec{q}s}} }	
	e^{-i\vec{q}\cdot\vec{r}_i}
% %
% \\
% %
%     u_{i\alpha} = &
%     \sqrt{ \frac{\hbar}{2N} }
%     \sum_{\vec{q}s}
% 	\frac{\epsilon_{\vec{q}s}^{i\alpha}}{\sqrt{m_i \omega_{\vec{q}s}}}
% 	e^{i\vec{q}\cdot\vec{r}_i} 
% 	(a_{\vec{q}s} + a_{\bar{\vec{q}}s}^\dagger)
% %
% \\
% %
%     p_{i\alpha} = &
%     -i \sqrt{ \frac{\hbar}{2N} }
%     \sum_{\vec{q}s}
% 	\epsilon_{\vec{q}s}^{i\alpha}
% 	\sqrt{m_i \omega_{\vec{q}s}}
% 	e^{i\vec{q}\cdot\vec{r}_i} 
% 	(a_{\vec{q}s} - a_{\bar{\vec{q}}s}^\dagger)
\end{align}
Here $\epsilon$ are phonon eigenvectors ($\vec{\epsilon}_i \cdot \vec{\epsilon}_j = \delta_{ij}$), $m$ atomic masses and $\omega$ frequencies. The vector $\vec{r}$ is a lattice vector, and $u$ and $p$ are the position and momentum operators. We use  the notation $\bar{\vec{q}}=-\vec{q}$. 
We introduce scaled eigenvectors (to simplify notation) via
\begin{equation}
	\upsilon_{\vec{q}s}^{i\alpha} = \sqrt{ \frac{\hbar}{2 m_i \omega_{\vec{q}s} } } \epsilon_{\vec{q}s}^{i\alpha}
\end{equation}
% %
% \\
% %
% 	a_{\vec{q}s} = & \frac{1}{\hbar\sqrt{N}}
% 	\sum_{i\alpha} \upsilon_{\vec{q}s}^{i\alpha}
% 	\left( m_i \omega_{\vec{q}s} u_{i\alpha}-i p_{i\alpha} \right) 
% 	e^{-i\vec{q}\cdot\vec{r}_i}
% %
% \\
% %
%     u_{i\alpha} = & \frac{1}{\sqrt{N}} 
% 	\sum_{\vec{q}s}
% 	\upsilon_{\vec{q}s}^{i\alpha}
% 	A_{\vec{q}s}
% 	e^{i\vec{q}\cdot\vec{r}_i} \,,\quad A_{\vec{q}s} = a_{\vec{q}s} + a{^\dagger}_{\bar{\vec{q}}s}
% %
% \\
% %
%     %p_{i\alpha} = & 
% 	a_{\vec{q}s} = & \frac{1}{\hbar\sqrt{N}}
% 	\sum_{i\alpha} \upsilon_{\vec{q}s}^{i\alpha}
% 	\left( m_i \omega_{\vec{q}s} u_{i\alpha}-i p_{i\alpha} \right) 
% 	e^{-i\vec{q}\cdot\vec{r}_i}
    %
%     \frac{1}{\sqrt{N}} 
% 	\sum_{\vec{q}s}
% 	\upsilon_{\vec{q}s}^{i\alpha}
% 	A_{\vec{q}s}
% 	e^{i\vec{q}\cdot\vec{r}_i} \,,\quad A_{\vec{q}s} = a_{\vec{q}s} + a{^\dagger}_{\bar{\vec{q}}s}
%
In this notation the matrix elements pertaining to anharmonicity become
\begin{align}
    \Phi_{\lambda\lambda'\lambda''} & = 
    \frac{1}{3!}
	\sum_{ijk\alpha\beta\gamma}
	\upsilon_{\vec{q}s}^{i\alpha}
	\upsilon_{\vec{q}'s'}^{j\beta}
	\upsilon_{\vec{q}''s''}^{k\gamma}
	\Phi_{ijk}^{\alpha\beta\gamma}
	e^{-i 
	(\vec{q}\cdot\vec{r}_i+ \vec{q}'\cdot\vec{r}_j+\vec{q}''\cdot\vec{r}_k)}
\\
    \Phi_{\lambda\lambda'\lambda''\lambda'''} & = 
    \frac{1}{4!}
	\sum_{ijkl\alpha\beta\gamma\delta}
	\upsilon_{\vec{q}s}^{i\alpha}
	\upsilon_{\vec{q}'s'}^{j\beta}
	\upsilon_{\vec{q}''s''}^{k\gamma}
	\upsilon_{\vec{q}'''s'''}^{l\delta}	
	\Phi_{ijkl}^{\alpha\beta\gamma\delta}
	e^{-i
	(\vec{q}\cdot\vec{r}_i +
	\vec{q}'\cdot\vec{r}_j +
	\vec{q}''\cdot\vec{r}_k +
	\vec{q}'''\cdot\vec{r}_l)
	} 
\intertext{so that the anharmonic part of the Hamiltonian becomes}
	H_A & = \Phi_{\lambda\lambda'\lambda''}A_{\lambda}A_{\lambda'}A_{\lambda''}+
	\Phi_{\lambda\lambda'\lambda''\lambda'''}A_{\lambda}A_{\lambda'}A_{\lambda''}A_{\lambda'''} + \cdots.	
\end{align}
In an analogous way we define the coefficients of the expansion of the polarizability as
\begin{align}
    \chi^{\mu\nu}_{\lambda} & =
	\sum_{i\alpha}
	\upsilon_{\vec{q}s}^{i\alpha}
	\chi_{i}^{\mu\nu,\alpha}
	e^{-i\vec{q}\cdot\vec{r}_i}
	\Delta_{\vec{q}} 
	=
	\sum_{i\alpha}
	\upsilon_{\vec{\Gamma}s}^{i\alpha}
	\chi_{i}^{\mu\nu,\alpha}
	=
	\chi^{\mu\nu}(s)
\\
\chi^{\mu\nu}_{\lambda\lambda'} & =
    \frac{1}{2!}
	\sum_{ij\alpha\beta}
	\upsilon_{\vec{q}s}^{i\alpha}
	\upsilon_{\vec{q}'s'}^{j\beta}
	\chi_{ij}^{\mu\nu,\alpha\beta}
	e^{-i\vec{q}\cdot\vec{r}_i}
	e^{-i\vec{q}'\cdot\vec{r}_j}
	\Delta_{\vec{q}\vec{q}'}
\\
	& =	
	\frac{1}{2!}
	\sum_{ij\alpha\beta}
	\upsilon_{\vec{q}s}^{i\alpha}
	(\upsilon_{\vec{q}s'}^{j\beta})^{\dagger}
	\chi_{ij}^{\mu\nu,\alpha\beta}
	e^{i\vec{q}\cdot(\vec{r}_j-\vec{r}_i)}
\\
	& =
	\frac{1}{2!}
	\sum_{ij\alpha\beta}
	(\upsilon_{\vec{q}s}^{i\alpha})^{\dagger}
	\upsilon_{\vec{q}s'}^{j\beta}
	\chi_{ij}^{\mu\nu,\alpha\beta}
	e^{-i\vec{q}\cdot(\vec{r}_j-\vec{r}_i)} = 
	\chi^{\mu\nu}(\vec{q},s,s') = 
	\chi^{\mu\nu}(\bar{\vec{q}},s,s')^\dagger =
	\chi^{\mu\nu}(\bar{\vec{q}},s',s)
\\
\chi^{\mu\nu}_{\lambda\lambda'\lambda''} & =
    \frac{1}{3!}
    \sum_{ijk\alpha\beta\gamma}
	\upsilon_{\vec{q}s}^{i\alpha}
	\upsilon_{\vec{q}'s'}^{j\beta}
	\upsilon_{\vec{q}'s'}^{k\gamma}
	\chi_{ijk}^{\mu\nu,\alpha\beta\gamma}
	e^{-i
	(\vec{q}\cdot\vec{r}_i+
	\vec{q}'\cdot\vec{r}_j+
	\vec{q}''\cdot\vec{r}_k)
	}
\intertext{Such that}
\label{eq:polarizabilityexpansion}
\chi^{\mu\nu} & = \chi^{\mu\nu}_0 + \chi^{\mu\nu}_{\lambda} A_{\lambda}
+ \chi^{\mu\nu}_{\lambda\lambda'} 
A_{\lambda}A_{\lambda'}
+ \chi^{\mu\nu}_{\lambda\lambda'\lambda''} 
A_{\lambda}A_{\lambda'}A_{\lambda''} + \cdots.
\end{align}

\vspace{5pt}

\subsection{Raman scattering in terms of the anharmonic Green's function}
If we start from the polarizability-polarizability correlation function and insert the expansion of the polarizability in terms of phonon coordinates, Eq.~\eqref{eq:polarizabilityexpansion}, we get
\begin{equation}
\label{eq:PP_expansion}
\begin{split}
    \left\langle \chi(t)\chi(0) \right\rangle = &
    \vec{\chi}_0 \otimes \vec{\chi}_0 + 
    \vec{\chi}_{\lambda} \otimes \vec{\chi}_{\lambda'} i\overset{>}{G}(A_{\lambda},A_{\lambda'})
+ \\
    + &
    \vec{\chi}_{\lambda\lambda'}
    \otimes
    \vec{\chi}_{\lambda''\lambda'''}
    i\overset{>}{G}(A_{\lambda}A_{\lambda'},A_{\lambda''}A_{\lambda'''})
+ \\    
     + &
    \vec{\chi}_{\lambda} \otimes \vec{\chi}_{\lambda'\lambda''}
    i\overset{>}{G}(A_{\lambda},A_{\lambda'}A_{\lambda''})
+
    \vec{\chi}_{\lambda\lambda'}
    \otimes
    \vec{\chi}_{\lambda''}
    i\overset{>}{G}(A_{\lambda}A_{\lambda'},A_{\lambda''})
+ \\
    + &
\vec{\chi}_{\lambda} \otimes \vec{\chi}_{\lambda'\lambda''\lambda'''}
    i\overset{>}{G}(A_{\lambda},A_{\lambda'}A_{\lambda''}A_{\lambda'''})
+
    \vec{\chi}_{\lambda\lambda'\lambda''}
    \otimes
    \vec{\chi}_{\lambda'''}
    i\overset{>}{G}(A_{\lambda}A_{\lambda'}A_{\lambda''},A_{\lambda'''}),
\end{split}
\end{equation}
where $\overset{>}{G}(A_{\lambda},A_{\lambda'}^{\dagger},\tau)\equiv-i\langle A_{\lambda}(\tau),A_{\lambda'}(0)\rangle$, and we have omitted terms of $A^4$ or higher.
The physical quantity of interest here is the spectral function, $J_{\lambda\lambda^{'}}(\Omega)$, which may also be written in terms of the retarded Green's function, $G^{R}_{\lambda\lambda'}(\Omega,T)= \mathrm{-}i\int\theta(\tau)\left<\left[A_{\lambda}(\tau),A^{\dagger}_{\lambda'}(0)\right]\right>_{T}
     e^{\mathrm{-}i\Omega t} \text{d}t,$:
\begin{equation}
\label{eq:spectral}
    J_{\lambda\lambda^{'}}(\Omega) = 
	- \frac{1}{\pi} \Imm\\\left\{ G^{R}_{\lambda\lambda^{'}}(\Omega) \right\} =	\frac{i}{(n(\Omega,T)+1)}\overset{>}{G}_{\lambda\lambda^{'}} (\Omega),
\end{equation}
where $n(\Omega,T)$ is the Bose-Einstein occupation number for energy $\hbar\Omega$ at temperature $T$, and with the retarded Green's function conveniently given in terms of the sum of a diagonal harmonic part, $g^0_{\lambda\lambda'}(\Omega)$, and the self-energy term $\Sigma_{\lambda\lambda'}(\Omega)$:
\begin{equation}
    G^{R}(\Omega)^{-1} = g^0(\Omega)^{-1} + \Sigma(\Omega).
\end{equation}    
Following the work of others~\cite{Leibfried1961,Cowley1963,wallace1998thermodynamics,Semwal1972}, we have previously discussed in detail the solution for the retarded Green's function of an anharmonic system $H_A$~\cite{Benshalom2022}.
In the case of third-order anharmonicity, the self-energy is given by
\begin{equation}
    \Sigma_{\lambda\lambda'}(\Omega) = -18 \sum_{\vec{q}_1\vec{q}_2 s_1 s_2}
    \Phi^{\lambda s_1s_2}_{\vec{q}\bar{\vec{q}}_1\bar{\vec{q}}_2}
    \Phi^{\lambda' s_1s_2}_{\bar{\vec{q}} \vec{q}_1 \vec{q}_2}
    S(s_1,s_2,\Omega)
\end{equation}
and
\begin{equation}
\begin{split}
    S(s_a,s_b,\Omega) = 
	(n_{a}+n_{b}+1) &
	\left[
	\frac{1}{(\omega_{a}+\omega_{b}-\Omega)_p} -
	\frac{1}{(\omega_{a}+\omega_{b}+\Omega)_p}
	\right]\\
	+
	 & (n_{a}-n_{b})
	\left[
	\frac{1}{(\omega_{b}-\omega_{a}+\Omega)_p}-
	\frac{1}{(\omega_{b}-\omega_{a}-\Omega)_p}
	\right].
\end{split}
\end{equation}

The constant (zeroth) term in Eq.~\eqref{eq:PP_expansion} does not contribute to the Raman spectrum. 
Using the spectral function~\eqref{eq:spectral}, the first term becomes 
\begin{equation}
\label{eq:SMramanI}
    I^{\textrm{I}} =
    \vec{\chi}_{\lambda} \otimes \vec{\chi}_{\lambda'}
    \int
    i \overset{>}{G}(A_{\lambda}(t),A_{\lambda'}(0))
    e^{-i\Omega t} dt
    =
    \vec{\chi}_{\lambda} \otimes \vec{\chi}_{\lambda'}
    (n(\Omega)+1)
    J_{\lambda\lambda'}(\Omega).
\end{equation}
This is sufficient to understand the origin of the polarization dependence presented in this study.
Higher terms in expansion~\eqref{eq:PP_expansion} are usually smaller, but become significant in Raman inactive structures, where the derivatives in Eq.~\eqref{eq:SMramanI} identically vanish.
This will happen for any crystal where all atomic sites constitute a center of inversion. 
See \cite{Benshalom2022} for a detailed discussion for the case of rock-salt structure, as well as how higher-order terms might play a role in finite temperatures.

\subsection{Light scattering in harmonic systems}
\label{har_vs_anh}

In harmonic systems all inter-atomic interactions are reduced to terms quadratic in their displacement from equilibrium in the potential expansion~\cite{Dove2003}. 
These produce ideal, non-interacting normal mode solutions, $G_{\lambda\lambda'}(\Omega)=g^0_{\lambda\lambda}(\Omega)\delta_{\lambda\lambda'}$~\cite{Cowley1963}.
Using the harmonic spectral function~\cite{Kwok1968}, the Stokes component in Eq.~\eqref{eq:SMramanI} becomes
\begin{equation}
\label{eq:harmonic_I}
    I(\Omega,T)=\sum_{\lambda}\vec{\chi}^{*}_\lambda\otimes\vec{\chi}_{\lambda}
    \frac{\left[n(\Omega,T)+1\right]}{2\omega_\lambda}
    \delta(\Omega-\omega_\lambda).
\end{equation}

The inadequacy of a purely harmonic treatment in light scattering is already partially recognized in most Raman studies, at least implicitly.
As Eq.~\eqref{eq:harmonic_I} shows, the frequency dependence of a truly harmonic system should be a delta-function. 
In practice, a Lorentzian-like line shape is commonly used, and explained by the same spectral function $J(\Omega)$, or more explicitly $\left(\Imm\\\{\Sigma_{\lambda\lambda}\}\right)^{-1}$, commonly known as the phonon lifetime~\cite{Maradudin1962}.
In addition, thermal expansion is sometimes incorporated through a quasi-harmonic treatment where the normal modes become volume dependent~\cite{Fultz2010}.

If only diagonal terms are considered (note that the self-energy need not vanish for this), the generalized fourth-rank expression of Eq.~\eqref{eq:fit} reduces back to the temperature \emph{independent} PO response (Eq.~(1) in main text), and all PO patterns may be explained in terms of a single second-rank tensor $\mathcal{R}$.
This is demonstrated by plugging Eq.~\eqref{eq:harmonic_I} into Eq.~\eqref{eq:fit}.
The incident field, $\vec{E}_i$, observed scattering polarization, $\vec{n}$, and second-rank tensor, $\vec{\chi}_{\lambda}$ are defined as:
\begin{equation}
  \vec{E}_{i}=\left|\vec{E}_i\right|\begin{pmatrix}e_{x}\\
e_{y}\\
e_{z}
\end{pmatrix};\,\,\,\,\,\,\,\,\,\,\,\vec{n}\equiv\hat{\vec{e}}_{s}=\begin{pmatrix}n_{x}\\
n_{y}\\
n_{z}
\end{pmatrix};\,\,\,\,\,\,\,\,\,\,\,\mathcal{R}\equiv\vec{\chi}_{\lambda}=\left(\begin{array}{ccc}
\chi_{xx} & \chi_{xy} & \chi_{xz}\\
\chi_{yx} & \chi_{yy} & \chi_{yz}\\
\chi_{zx} & \chi_{zy} & \chi_{zz}
\end{array}\right).  
\end{equation}
The polarization dependence of the scattering cross-section for a single mode $\lambda$ becomes:
\begin{align*}
    \sigma\propto & \sum_{\mu\nu\xi\rho}n_{\mu}n_{\xi}I_{\mu\nu,\xi\rho}E_{\nu}E_{\rho}
    \\
    = & \left[n(\omega_\lambda,T)+1\right]\frac{\left|E_i\right|^2}{2\omega_\lambda}
    \sum_{\mu\nu\xi\rho}n_{\mu}n_{\xi}\left(\vec{\chi}^{*}_\lambda\otimes\vec{\chi}_{\lambda}\right)_{\mu\nu\xi\rho}e_{\nu}e_{\rho}
    \\
    \propto & \sum_{\mu\nu\xi\rho}n_{\mu}n_{\xi} \chi^{*}_{\nu\mu}\chi_{\rho\xi} e_{\nu}e_{\rho}
    \\
    = & \left(\sum_{\nu\mu}\chi^{*}_{\nu\mu}n_{\mu}e_{\nu} \right) \left(\sum_{\rho\xi}\chi_{\rho\xi}n_{\xi}e_{\rho}\right)
    = \left(\sum_{\nu\mu}\chi_{\nu\mu}n_{\mu}e_{\nu} \right)^{*} \left(\sum_{\rho\xi}\chi_{\rho\xi}n_{\xi}e_{\rho}\right)
    \\
    = & \left|\sum_{i j}\chi_{i j}n_{j}e_{i}\right|^{2} 
    = \left|\sum_{i}e_{i}\sum_{j}\chi_{i j}n_{j}\right|^{2}
    \\
    = & \left|\hat{\vec{e}}_{i}\cdot\mathcal{R}\cdot\hat{\vec{e}}_{s}\right|^{2},
\end{align*}
which is exactly the original second-rank expression.

\clearpage

\printbibliography
%\bibliographystyle{unsrt}
%\bibliography{mylibrary}
\renewcommand{\thepage}{S\arabic{page}}  
\renewcommand{\thesection}{S\arabic{section}}   
\renewcommand{\thetable}{S\arabic{table}}   
\renewcommand{\thefigure}{S\arabic{figure}}